%% file: Main.tex
\documentclass[proof]{WileyASNA-v1}

\articletype{Article Type}%

\received{29 April 2021}
\revised{16 June 2021}
\accepted{11 August 2021}
\usepackage{comment}
\begin{document}

\title{Simulating starspot activity jitter for spectral types F--M: realistic estimates for a representative sample of known exoplanet hosts}

\author[1]{Stefano Bellotti*}

\author[2,3]{Heidi Korhonen}

\authormark{Bellotti \& Korhonen}

\address[1]{\orgdiv{Institut de Recherche en Astrophysique et Plan\'etologie, Universit\'e de Toulouse}, \orgname{CNRS, IRAP/UMR 5277}, \orgaddress{\state{14 avenue E. Belin, Toulouse, F-31400}, \country{France}}}

\address[2]{\orgdiv{European Southern Observatory (ESO)}, \orgaddress{\state{Alonso de Córdova 3107, Vitacura, Santiago}, \country{Chile}}}

\address[3]{\orgdiv{DARK, Niels Bohr Institute, University of Copenhagen}, \orgaddress{\state{Jagtvej 128, 2200 Copenhagen}, \country{Denmark}}}

\corres{*Stefano Bellotti \email{stefano.bellotti@irap.omp.eu}}


\abstract{Dark spots on the surface of active stars produce changes in the shapes of the spectral lines that mimic spurious Doppler shifts, compromising the detection of small planets by means of the radial velocity (RV) technique. Modelling the spot-driven RV variability (known as ``jitter'') and how it affects the RV data sets is therefore crucial to design efficient activity-filtering techniques and inform observing strategies. Here, we characterise starspots and simulate the radial velocity curves induced by them to determine typical jitter amplitudes for a representative sample of 15 known host stars spanning between F and M spectral type. We collect information on the $\log R'_{\mathrm{HK}}$ activity index from the literature for 205 stars and, due to a lack of data in the temperature range 4000-4500\,K, we measure it for ten stars using archival data. Additional stellar parameters required for the simulations are collected from the literature or constrained by observational data, in order to derive realistic estimates. Our results can be used as reference to determine typical peak-to-peak spot-induced RV jitter in the visible domain that can be expected when targeting host stars with different properties.}


\jnlcitation{\cname{%
\author{Bellotti S.} 
\author{\& Korhonen H.}} (\cyear{2021}), 
\ctitle{Title}, \cjournal{Astronomical Notes}, \cvol{2017;00:1--6}.}


\maketitle


\input{Sec1.tex}
\input{Sec2.tex}
\input{Sec3.tex}
\input{Sec4.tex}
\input{Sec5.tex}

\backmatter

\section*{Acknowledgments}
HK acknowledges a research grant from the Danish Augustinus Fonden. This research has made use of the SIMBAD database, operated at CDS, Strasbourg, France; the VizieR catalogue access tool, CDS, Strasbourg, France; the NASA Exoplanet Archive, which is operated by the California Institute of Technology, under contract with the National Aeronautics and Space Administration under the Exoplanet Exploration Program; Astropy, 12 a community-developed core Python package for Astronomy \citep{Astropy2013,Astropy2018}; Matplotlib: Visualization with Python \citep{Hunter2007}; SciPy \citep{Virtanen2020}. Based on observations collected at the European Southern Observatory under ESO programmes 082.C-0390(A), 086.D-0236(A), 185.D-0056(J), 090.C-0146(A), 089.C-0444(A), 072.C-0488(E), 072.B-0585(A), 094.C-0707(A), 0100.A-9006(A).







\appendix
\input{appendix.tex}

\bibliography{biblio}%


\end{document}

%% file: Sec1.tex
\section{Introduction}

The radial velocity (RV) technique is the second most prolific exoplanet hunting method, counting 21\% of the total confirmed discoveries\footnote{\href{exoplanet.eu}{exoplanet.eu}, December 2020.}. Central interest is given to the detection and characterization of habitable Earth-mass planets, which requires a precision on the order of m s$^{-1}$ or cm s$^{-1}$ for an M or G type star, respectively. Although instrumentation has witnessed significant advancements in this direction with ESPRESSO \citep{Pepe2013}, NEID \citep{Schwab2016} and EXPRES \citep{Jorgenson2016}, stellar activity represents a serious limitation, as the associated RV jitter can completely swamp the planetary signal by several orders of magnitude \citep[e.g.,][]{Saar1997, Meunier2010, Donati2016}. 

The activity-driven RV variability is caused by a diversity of phenomena taking place on different timescales (see \citealt{Meunier2021}): oscillations and granulation induce RV signals on the order of m s$^{-1}$ over minutes and hours \citep{Dumusque2011}, whereas faculae, spots, and magnetic cycles lead to signals of m s$^{-1}$-km s$^{-1}$ over days and years \citep{Hussain2002, Lanza2010}. The common denominator for most of these phenomena with the largest contributions resides in the magnetic field of the star \citep{Schrijver2008}.

Particular importance is given to faculae and spots, since their effect represent the main obstacle when searching for small planets. Faculae are responsible for the inhibition of convective motions, with consequent suppression of the blueshift arising from granulation \citep{Meunier2010,Haywood2014,Miklos2020}, while starspots distort the spectral line profiles when they cross the visible stellar disk \citep{Saar1997}, producing noise or mimicking radial velocity variations due to a planet and therefore resulting in erroneous detections \citep[e.g.,][]{Queloz2001,Huelamo2008,Huerta2008}.

Modelling the starspot signature affecting RV data sets is therefore key to improve the precision of planet searches and inform observing strategies to minimize the impact of activity. To this aim, a number of tools have been developed such as \textsc{soap} \citep{Boisse2012,Dumusque2014}, \textsc{deema} \citep{Korhonen2015} and \textsc{StarSim} \citep{Herrero2016}. 

In general, the impact of starspots on RV data sets and planet detectability has been simulated under different assumptions \citep{Meunier2021}. For instance, \citet{Desort2007} studied the case of a single spot on the surface of different F--K stars and with different active latitudes, inclinations and $v\sin i$; \citet{Lagrange2010} investigated planet detection limits for G type stars using Sun-like spot coverage; \citet{Santos2015} used an empirical model to reconstruct a synthetic solar sunspot cycle; \citet{Andersen2015} explored the M type regime and various activity levels, and \citet{Dumusque2016} included instrumental, stellar and planetary signals in simulated RV data sets for G--K stars to analyze different recovery techniques. 

These works studied the effects of different activity configurations or stellar parameters on the RV data sets. In comparison, we select a representative sample of known exoplanet hosts in the F--M spectral range, find or compute their stellar parameters, characterise their starspot properties, and simulate their spotted surface to obtain reasonable estimates of the spot-induced jitter. These can then be used to approximately predict peak-to-peak jitter values in the optical domain when targeting stars of analogous properties. We do not include faculae in our simulations because of their limited observational knowledge for other stars (e.g., filling factor). In addition, the facula-photosphere temperature contrast is predicted to be at most 3\% and 10\% of the spot-photosphere contrast for weak and strong magnetic fields, respectively \citep{Johnson2021}. This would imply that the jitter introduced by starspots would dominate over the contribution from the faculae.

The paper is structured as follows. The list of magnetic activity for known host stars and the selection of a representative sample is detailed in Section~\ref{sec:starsample}. We then describe the estimates of the stellar parameters required for the simulations in Section~\ref{sec:simparameters}. The parameters are used as input to produce synthetic spectra by means of the \textsc{deema} code \citep{Korhonen2015}. Finally, we compute the spot-induced RV curves in Section~\ref{sec:simulations} and discuss our results in Section~\ref{sec:results}.

%% file: Sec2.tex
\section{Host stars sample} \label{sec:starsample}

We compile a list of magnetic activity for known host stars between M and F type, the spectral range containing most of the planet hosts \footnote{\href{exoplanet.eu}{exoplanet.eu}}. The activity is quantified by the chromospheric activity index $\log R'_{\mathrm{HK}}$ representing the contribution of the CaII H \& K lines to the bolometric luminosity of the star and corrected for the photospheric term \citep{Noyes1984}. This indicator was introduced by \citet{Middelkoop1982} to substitute the colour dependent $S$ index, originally used in the Mount Wilson Observatory HK Project \citep{Vaughan1978} to characterize stellar activity. Since $\log R'_{\mathrm{HK}}$ correlates with the presence of spots on the stellar surface, i.e. sources of activity jitter, its temporal variation is often employed to discriminate between activity and genuine planetary signals \citep{Bonfils2007}.

Our list contains information on effective temperature ($T_{\mathrm{eff}}$), $B-V$ colour index, spectral type, and $\log R'_{\mathrm{HK}}$ index available either in the literature or on online databases. When the $S$ index is provided instead, we convert it to the $\log R'_{\mathrm{HK}}$ index with \citet{Rutten1984} formula.

For all F, G and some K stars, we extract data from the catalogue by \citet{Krejcova2012}, who investigated the influence of an exoplanet on the magnetic activity of its host star. For later spectral types, we find known host stars via exoplanet.eu \citep{exoeu} and exoplanet.org \citep{exoorg} and we look for magnetic activity information in the literature. Moreover, we use SIMBAD \citep{SIMBAD} to obtain $B-V$ colour indexes and as an auxiliary tool when the spectral type and evolutionary stage are uncertain. 

The list obtained from these sources shows a dearth of targets with $T_{\mathrm{eff}}$=4000--4500\,K and $\log R'_{\mathrm{HK}}<-5.0$ \citep{Saikia2018}. Therefore, we look for known host stars on exoplanet.eu in this temperature range, retrieve the corresponding spectra from the ESO archive, and measure the $\log R'_{\mathrm{HK}}$ index following \citet{Duncan1991}.

\subsection{Activity index measurements} \label{sec:Smeasures}
We add ten host stars to our list with $T_{\mathrm{eff}}$=4000--4500\,K and with available ESO-archive spectra. For each star, we select the spectrum with the highest signal-to-noise ratio  collected with high-resolution spectrographs. This results in three spectra from HARPS, five from FEROS and two from UVES. HARPS is the high accuracy radial velocity planets searcher installed on the ESO 3.6 m telescope at La Silla observatory in Chile. It is a fiber-fed cross-dispersed echelle spectrograph covering the 3800-6900 \r{A} spectral range with a resolving power of 115000 \citep{Mayor2003}. FEROS is a fiber-fed extended range optical spectrograph operating at ESO La Silla as well. Its resolving power is 48000 and it covers the wavelength range 3600-9200 \r{A} over 39 orders \citep{Kaufer1999}. UVES is the ultraviolet and visible echelle spectrograph of the VLT. Its resolving power is about 40000 when a 1 arcsec slit is used. The maximum (two-pixel) resolution is 80000 or 110000 in the blue and the red arm respectively \citep{Dekker2000}.

Estimating the activity index consists of 1) measuring the $S$ index, 2) calibrating it to the Mount Wilson scale and 3) converting it to $\log R'_{\mathrm{HK}}$. 

\begin{figure}[t]
    \centerline{\includegraphics[width=\columnwidth]{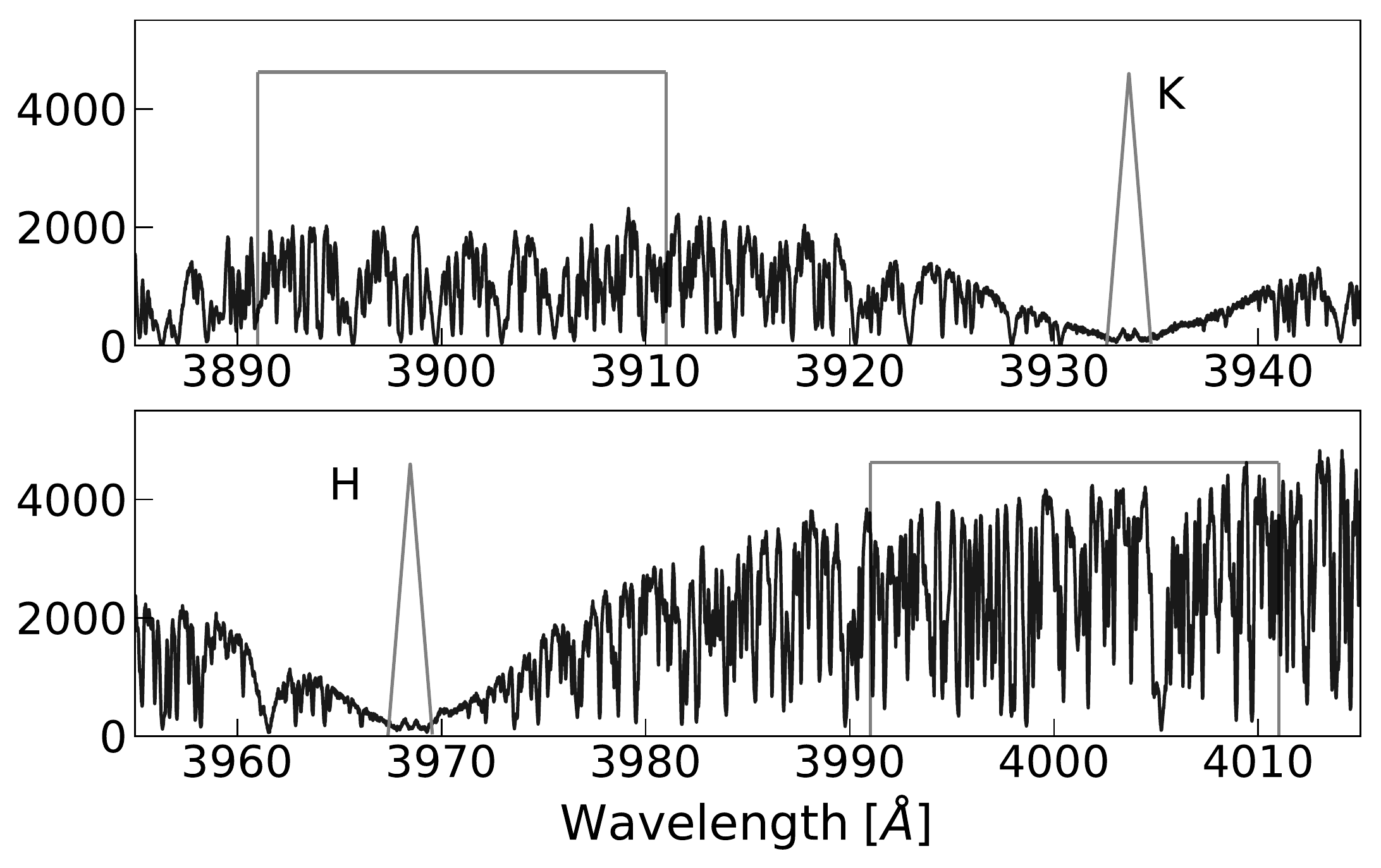}}
    \caption{Passbands in the definition of the $S$ index. Shown is a HARPS spectrum for HD\,47536. Top: V and K passbands. Bottom: H and R passbands. The vertical axis represents the number of counts.\label{fig:bands}}
\end{figure}

1) The wavelength axis in the spectra is shifted to account for the radial velocity of the star. We then reproduce the measurements of the $S$ index following \citet{Duncan1991}: we define two triangular passbands with FWHM = 1.09 {\AA} centered on the cores of the K (3933.661 \AA) and H (3968.470 \AA) lines, and two 20 {\AA} wide rectangular passbands centered on 3901 {\AA} (V band) and 4001 {\AA} (R band), respectively (Fig.~\ref{fig:bands}). The $S$ index is defined as
\begin{equation}
    S=\alpha\frac{N_\mathrm{H}+N_\mathrm{K}}{N_\mathrm{R}+N_\mathrm{V}}
    \label{eq:Sindex}
\end{equation}
where $N_i$ are the counts in the corresponding i-th passband and $\alpha$ is a calibration constant \citep{Vaughan1978}. $\alpha$ is used to relate the values obtained with HKP-2 to the HKP-1 scale (the first two spectrometers of the Mount Wilson project); it is set to 2.4 and, in turn, it is multiplied by 8 which is a correction factor due to the longer exposure times of the V and R passbands of the HKP-2 instrument relative to HKP-1.

\begin{figure}[t]
    \centerline{\includegraphics[width=\columnwidth]{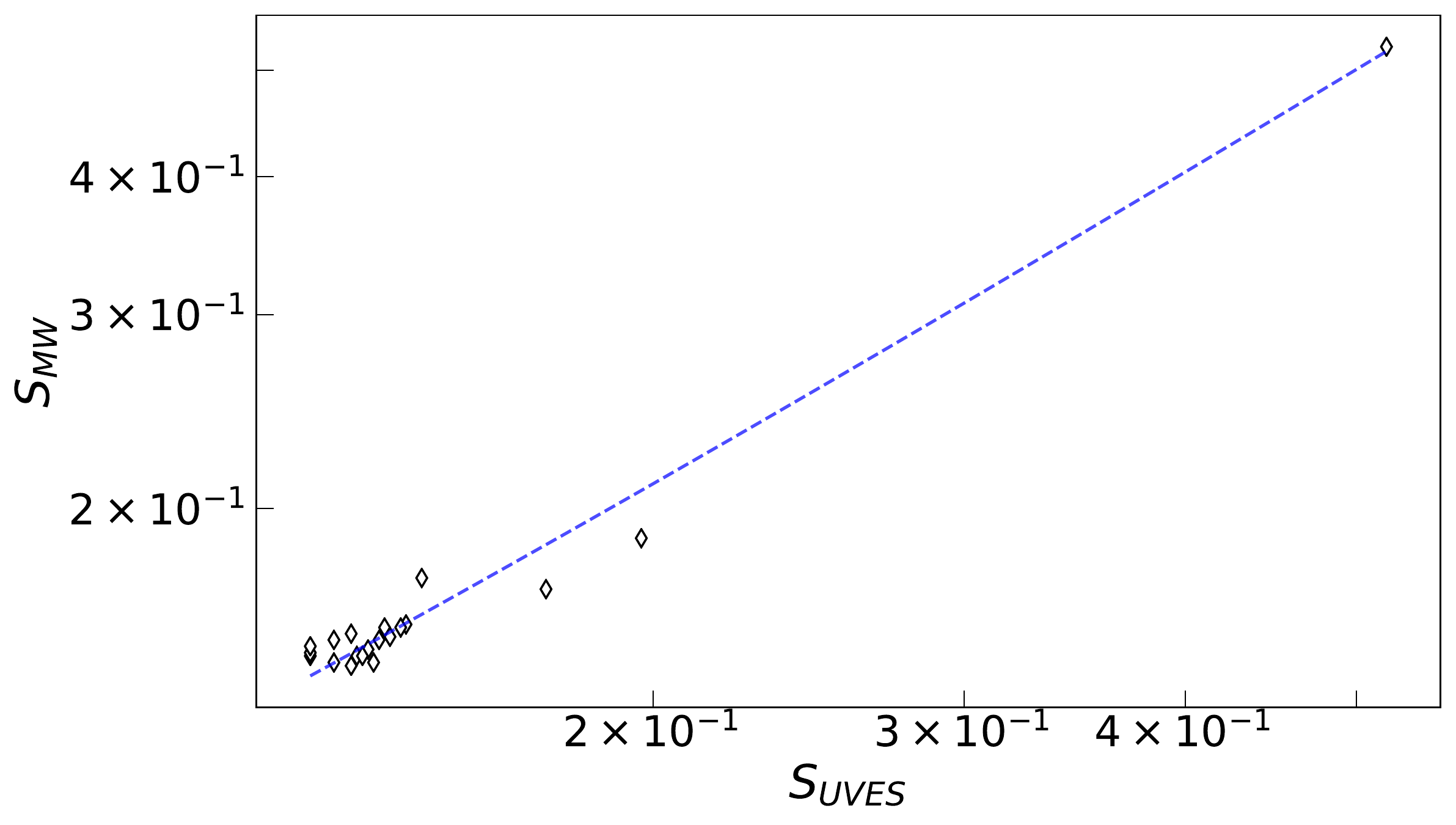}}
    \caption{Calibration of our $S$ index values measured with UVES and literature values \citep{Wright2004}. The dashed blue line represents the least-square linear fit between the data sets. The RMS of the residuals is 0.0027, plausibly due to measurement errors, intrinsic variability of the chromosphere, or the effect of long-term activity trends, given that we measure indices using observations taken at least ten years apart \citep{Wright2004}.
    \label{fig:Sfit}}
\end{figure}

2) For HARPS and FEROS, the calibration equations from the instrument scale to the Mount Wilson scale are given by \citet{Saikia2018} and \citet{Jeffers2018}, respectively. They are,
\begin{gather}
    S_{\mathrm{MW}}=1.1159\cdot S_{\mathrm{HARPS}}+0.0343\\
    S_{\mathrm{MW}}=1.6880\cdot S_{\mathrm{FEROS}}+0.0600
\end{gather}
For UVES, the calibration equation is not present in the literature; we compute it taking the 20 stars we have in common with \citet{Wright2004} and comparing their $S$ index values (already converted to Mount Wilson scale) with ours. We perform a linear fit of the two $S$ index data sets (Fig.~\ref{fig:Sfit}), resulting in
\begin{equation}
    S_{\mathrm{MW}}=0.967\;(\pm0.022)\cdot S_{\mathrm{UVES}}+0.017\;(\pm 0.004)
    \label{eq:linearfit}
\end{equation}

3) To convert the $S$ indexes in $\log R'_{\mathrm{HK}}$, we use the colour-dependent equation in \citet{Rutten1984}. The formula has a broad empirical validity in the $B-V$ range and distinguishes between main sequence and giant stars.

The measured activity indexes are summarised in Table~\ref{tab:RHKresults}. We extract the uncertainties on $B-V$ from the works of \citet{Hog2000} and \citet{Zacharias2013} or SIMBAD, when available, whereas we infer the uncertainties on $S$ index and on $\log R'_{\mathrm{HK}}$ by error propagation. Contrarily to UVES, the uncertainty on the flux counts is not provided by HARPS and FEROS pipelines. In these cases, we infer the signal-to-noise ratio from a continuum region of the spectrum close to the H \& K lines and use its reciprocal as an estimate of the error for all the bands that enter Eq.~(\ref{eq:Sindex}).

The fact that some measurements present wide error bars can be attributed to: the choice of continuum region over which to estimate the signal-to-noise ratio, the propagation of large ($\sim$0.1) $B-V$ errors, or a low signal-to-noise ratio of the spectrum. Typically the spectra have signal-to-noise ratio of at least 50, but in the case of HAT-P-54 only $\sim$5.

\begin{center}
\begin{table*}[t]
    \caption{Activity information for known host stars with $T_{\mathrm{eff}}$=4000--4500\,K. The $\log R'_{\mathrm{HK}}$ index is either converted directly from the literature $S$ index (first three stars) or measured by us. HD\,208527 is taken from \citet{Wright2004}, whereas HD\,73108 and HD\,21552 from \citet{Duncan1991}. \label{tab:RHKresults}}
    \centering
    \begin{tabular*}{42pc}{@{\extracolsep\fill}lclccl@{\extracolsep\fill}}
        \toprule
        \textbf{ID} & \textbf{Instrument} & $\textbf{B}-\textbf{V}$ & \textbf{Spectral type} & $\textbf{S}$ & \textbf{log} $\textbf{R}'_{\mathrm{\textbf{HK}}}$\\
        \midrule
        HD\,208527 & HIRES & $1.698\pm 0.114$ & K III & $0.288\pm0.0084$ & $-5.840\pm0.050$\\
        HD\,73108 & HKP-2 & 1.170 & K III & $0.124\pm0.0000$ & $-5.360$\\
        HD\,21552 & HKP-2 & 1.350 & K III & $0.158\pm0.0024$ & $-5.500$\\
        HD\,47536 & HARPS & $1.172\pm0.013$ & K III & $0.142\pm0.001$ & $-5.310\pm0.020$\\
        HD\,110014 & HARPS & $1.243\pm0.017$ & K III & $0.127\pm0.001$ & $-5.445\pm0.025$\\
        HIP\,90979 & HARPS & $1.220\pm 0.112$ & K V & $0.798\pm0.002$ & $-4.730\pm0.190$\\
        HD\,66141 & FEROS & 1.250 & K III & $0.230\pm 0.005$ & $-5.200$\\
        $\gamma$\,Leo A & FEROS & $1.420\pm 0.014$ & K III & $0.195\pm0.003$ & $-5.520\pm0.020$\\
        BD+20\,2457 & FEROS & $1.250\pm0.085$ & K II & $0.290\pm0.012$ & $-5.090\pm0.120$\\
        BD+20\,274 & FEROS & $1.350\pm 0.076$ & K III & $0.261\pm0.070$ & $-5.290\pm0.220$\\
        17\,Sco & FEROS & $1.376\pm 0.017$ & K III & $0.175\pm0.007$ & $-5.500\pm0.030$\\
        HAT-P-54 & UVES & $1.330\pm 0.092$ & K V & $1.373\pm0.002$ & $-4.680\pm0.160$\\
        $\varepsilon$\,CrB & UVES & $1.230\pm 0.010$ & K III & $0.1200\pm0.0001$ & $-5.470\pm0.150$\\
        \bottomrule
    \end{tabular*}
\end{table*}
\end{center}

\begin{figure}[t]
    \centerline{\includegraphics[width=\columnwidth]{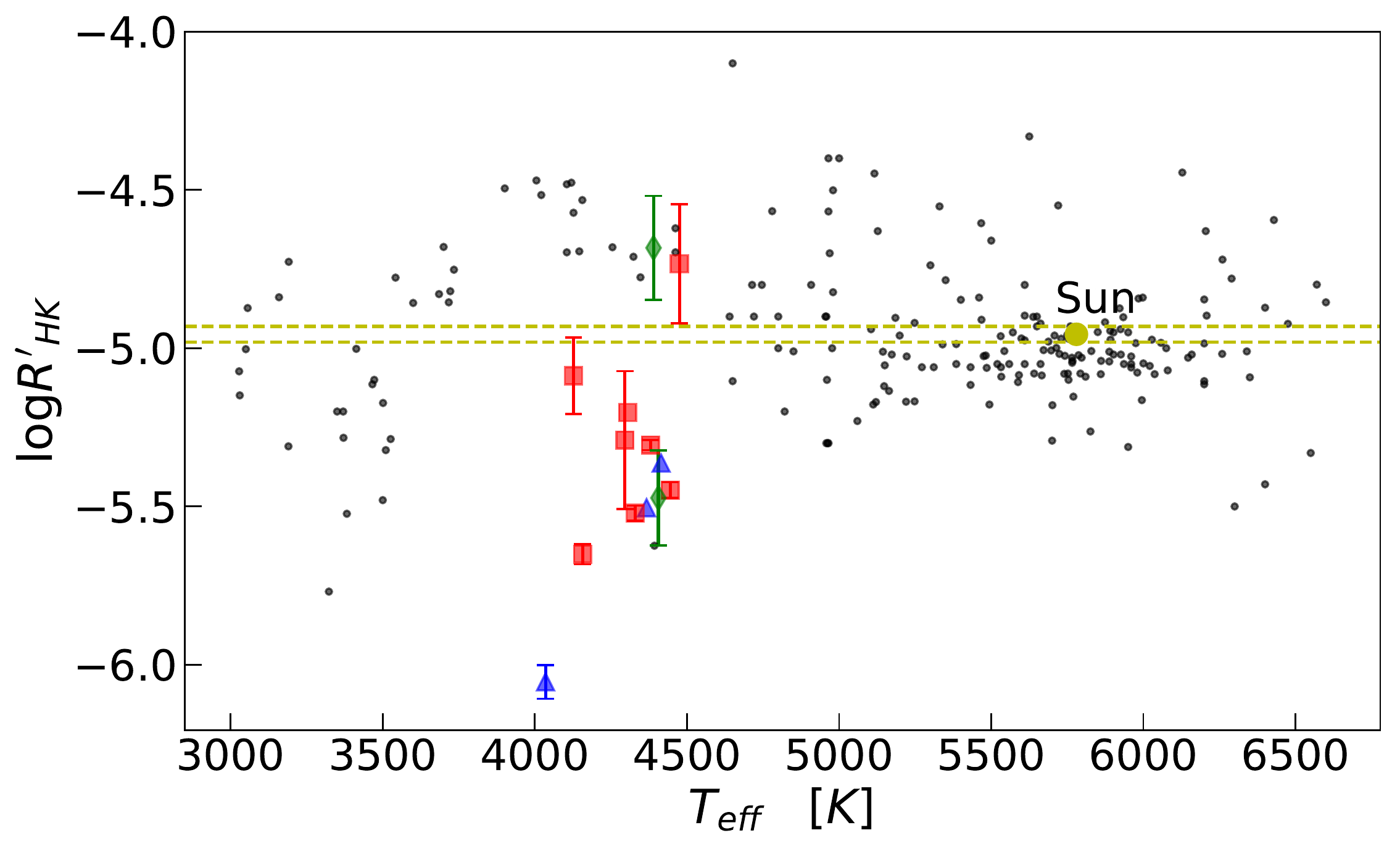}}
    \caption{Complete list of magnetic activity for the host stars. Shown are the chromospheric activity values from the literature (black dots) or measured by us: red squares for HARPS and FEROS and green diamonds for UVES. The blue triangles represent stars for which the $S$ index is in the literature instead of $\log R'_{\mathrm{HK}}$. The Sun is added for comparison purposes, with two dashed horizontal lines representing the activity level during solar maximum and minimum in cycle 24, respectively \citep{Egeland2017}.
    \label{fig:catalogue}}
\end{figure}

\begin{figure}[t]
    \centerline{\includegraphics[width=\columnwidth]{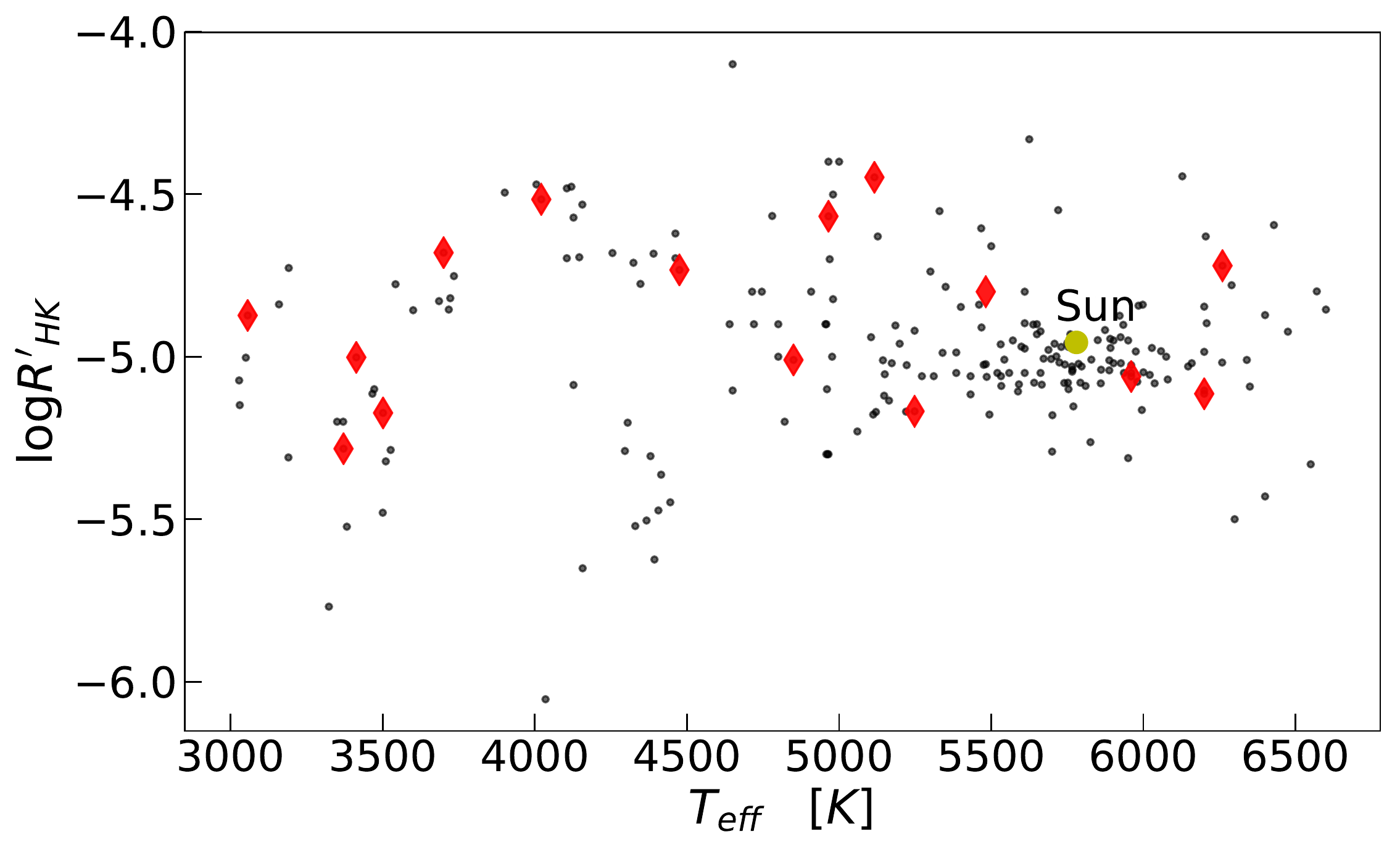}}
    \caption{Sample of 15 stars used for the simulations. It is representative of known planet hosts for various activity levels and over the M--F spectral types. 
    \label{fig:sample}}
\end{figure}

\subsection{Simulation sample} \label{sec:simsample}

Adding the newly measured activity indices to the literature values results in a magnetic activity list composed of 218 stars of which 28 are F type, 90 G type, 69 K type, and 31 M type. The complete list is illustrated in Fig.~\ref{fig:catalogue}.

To simulate a representative sample of typical planet host stars, we select main sequence stars that are homogeneously distributed in the temperature-activity parameter space. The region $T_{\mathrm{eff}}$=4000--4500\,K and $\log R'_{\mathrm{HK}} < -5.0$ contains only stars that have evolved off the main sequence (see Table~\ref{tab:RHKresults}), hence it is not represented.

We obtain a final simulation sample of 15 stars between 3000 K and 6500 K, with the chromospheric activity index spanning between -4.448 ($\varepsilon$ Eri) and -5.283 (GJ\,180), as shown in Fig.~\ref{fig:sample}.

%% file: Sec3.tex
\begin{figure*}[t]
    \centerline{\includegraphics[width=\textwidth]{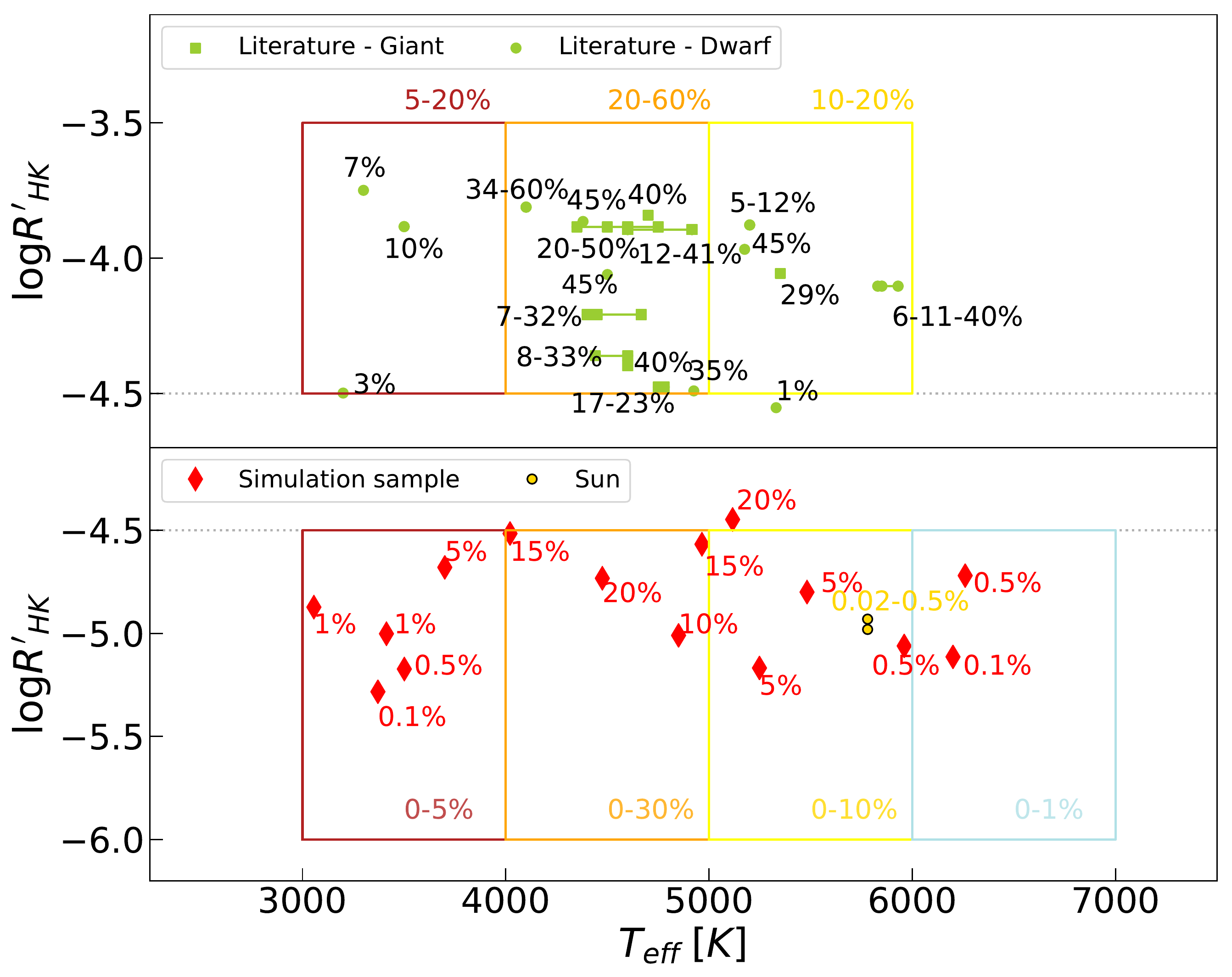}}
    \caption{
    Procedure adopted to constrain and estimate the spot filling factors for our simulation sample. Top: Known filling factor values. Observational data for giant (green squares) and main sequence (green circles) stars are shown. Multiple values for an individual star are connected, and both maximum and minimum are reported. The squared boxes represent the range of typical filling factor values over different spectral types, with associated values displayed on the top. Bottom: Our simulation sample. The squared boxes represent the extrapolated ranges of filling factors (given at the bottom of each box). These extrapolated regions are set to span between 0\%, at $\log R'_{\mathrm{HK}}$ = -6.0, and a value that matches the observational boxes at $\log R'_{\mathrm{HK}}$ = -4.5 \citep{Afram2015,SuarezMascareno2018}. The central extrapolated region (4000-5000 K) reaches 30\%, which is approximately the average of observational values close to $\log R'_{\mathrm{HK}}$ = -4.5. The simulation sample is shown (red diamonds) along with the estimates of the filling factor. These values are obtained considering that the filling factor decreases toward lower activity levels, and toward hotter and cooler stars than K type.} \label{fig:sample_ff}
\end{figure*}

\section{Simulation parameters} \label{sec:simparameters}
To generate synthetic stellar spectra, we require: stellar mass, photospheric temperature, spot temperature, spot filling factor, stellar inclination, stellar rotation period, active longitudes and latitudes. Stellar masses and temperatures are mainly retrieved from exoplanet.eu, while the remaining parameters are either computed using empirical relations or estimated based on data from previous works.

\begin{center}
\begin{table*}[t]
    \caption{Estimate properties of the simulated stars: chromospheric activity index, spectral type, stellar mass, effective temperature, spot temperature, filling factor, rotation period, projected rotational velocity, and active latitude band. All spectral types, effective temperatures and masses are taken from exoplanet.eu or from an otherwise specified reference. \label{tab:sim_params}}
    \centering
    \begin{tabular*}{42pc}{@{\extracolsep\fill}lllcllrccc@{\extracolsep\fill}}
        \toprule
        \textbf{ID} & \textbf{log} $\textbf{R}'_{\mathrm{\textbf{HK}}}$ & \textbf{Spectral type} & $\textbf{M}_{*}\;[\textbf{M}_{\bigodot}]$ & $\textbf{T}_{\mathrm{\textbf{eff}}}\;[\mathrm{\textbf{K}}]$ & $\textbf{T}_{\mathrm{\textbf{spot}}}\;[\mathrm{\textbf{K}}]$ & \textbf{ff [\%]} & $\textbf{P}_{\mathrm{\textbf{rot}}}$ [\textbf{d}] & \textbf{vsini} [\textbf{km s$^{-1}$}] & $\pm\Delta \textbf{l}\;[^\circ]$\\
        \midrule
        YZ\,Cet & -4.873\tnote{$^b$} & M4.5 V & 0.130 & 3056 & 2540 & 1 & 26 & 0.3 & $\pm(35-60)$\\
        GJ\,180 & -5.283\tnote{$^b$} & M2 V & 0.430 & 3371 & 2790 & 0.1 & 54 & 0.4 & $\pm(30-55)$\\
        GJ\,687 & -5.002\tnote{$^c$} & M3.5 V & 0.413 & 3413 & 2820 & 1 & 32 & 0.6 & $\pm(35-60)$\\
        GJ\,3634 & -5.173\tnote{$^b$} & M2.5 V\tnote{$^f$} & 0.450 & 3501\tnote{$^f$} & 2890 & 0.5 & 45 & 0.5 &$\pm(30-55)$\\
        GJ\,649  & -4.680\tnote{$^c$} & M1.5 V\tnote{$^f$} & 0.540 & 3700 & 3040 & 5 & 18 & 2.9 & $\pm(20-65)$\\
        HIP\,54373  & -4.516\tnote{$^c$} & K5 V & 0.570 & 4021 & 3280 & 15 & 22 & 1.3 & $\pm(20-65)$\\
        HIP\,90979 & -4.733\tnote{$^a$} & K7 V & 0.571 & 4475 & 3600 & 20 & 37 & 0.8 & $\pm(15-60)$\\
        HD\,156668 & -5.010\tnote{$^c$} & K3 V & 0.772 & 4850 & 3860 & 10 & 49 & 0.8 & $\pm(15-60)$\\
        HD\,192263 & -4.568\tnote{$^c$} & K2 V & 0.804 & 4965 & 3940 & 15 & 25 & 1.6 & $\pm(10-60)$\\
        $\varepsilon$\,Eri & -4.448\tnote{$^d$} & K2 V & 0.830 & 5116 & 4035 & 20 & 13 & 3.1 & $\pm(20-70)$\\
        HD\,11964  & -5.168\tnote{$^d$} & G5 & 1.125 & 5248 & 4120 & 5 & 50 & 1.1 & $\pm(0-65)$\\
        HD\,49674 & -4.800\tnote{$^d$} & G5 V & 1.015 & 5482 & 4270 & 5 & 27 & 2.0 & $\pm(0-65)$\\
        HAT-P-5 & -5.061 & G1 V\tnote{$^e$} & 1.163 & 5960 & 4560 & 0.5 & 51 & 1.0 & $\pm(0-65)$\\
        WASP-1 & -5.114\tnote{$^d$} & F7 V & 1.240 & 6200 & 4700 & 0.1 & 40 & 1.3 & $\pm(0-65)$\\
        HD\,179949 & -4.720\tnote{$^d$} & F8 V & 1.181 & 6260 & 4740 & 0.5 & 8 & 6.6 & $\pm(10-65)$\\
        \bottomrule
    \end{tabular*}
    \begin{tablenotes}
    \item Sources: $^a$ This work, $^b$ \citet{Astudillo2017}, $^c$ \citet{Saikia2018}, $^d$ \citet{Krejcova2012}, $^e$ \citet{Faedi2013}, $^f$ \citet{Houdebine2019}.
    \end{tablenotes}
\end{table*}
\end{center}

\subsection{Spot temperature and filling factor}\label{sec:Temp_ff}

Different methods can be applied to derive starspot properties, e.g. Doppler and Zeeman-Doppler Imaging, light curve modelling, molecular bands modelling, atomic line-depth ratios, and each with a different level of accuracy. Combining the results of these methods, \citet{Berdyugina2005} argued that starspots can be 500 to 2000 K cooler than the quiet photosphere and cover more than 30\% of the stellar surface. The reported empirical fits have been recently implemented by \citet{Herbst2021} with an extension of the stellar sample over which the fit is computed.

For the spot temperature contrast, we use the stellar-temperature dependent formula (Eq. 6 in \citealt{Herbst2021}) and we obtain values between 500 and 1500 K for M--F type stars, in agreement with \citet{Berdyugina2005} and \citet{Andersen2015}.

For the filling factor, the lack of an analytical expression relating it to the chromospheric activity level makes the estimation more complicated. There are empirical fits for G type stars to determine the starspot size based on stellar radius, stellar temperature and photometric variations \citep{Maehara2017}, but they are valid when large spots groups and long spot lifetimes can be both assumed. Knowing that starspots lifetimes span between few and several stellar rotations \citep{Namekata2019} and that our simulations include M type stars, which can be covered with small spots \citep{Jackson2012}, we do not use these fits to predict the filling factor for our simulation stars.

Fig.~\ref{fig:sample_ff} clarifies the adopted procedure instead. For activity levels above $\log R'_{\mathrm{HK}} =$ -4.5, we identify three regions representing typical filling factors according to spectral type \citep{Afram2015,SuarezMascareno2018}. To support the values in these regions, we also include observational filling factors for both active giants and main sequence stars (reported in Table~\ref{tab:ff_data}). The $\log R'_{\mathrm{HK}}$ value of -4.5 is reasonably consistent with the boundary between active and inactive stars in \citet{Henry1996}. 

At this point, we 1) extrapolate the three filling factor regions towards lower activity levels considering that the number of spots decreases with decreasing activity \citep{Balmaceda2009} and 2) measure individual filling factors knowing that this quantity peaks for K stars (at $\sim$ 4500\;K) and decreases quadratically towards both cooler and hotter stars (Fig. 10 in \citealt{Berdyugina2005}). The latter feature is also used to obtain a filling factor region for F stars. 

This filling factor estimate comes with important caveats. Because of the paucity of the data or contradicting values (e.g. EK Dra), some assumptions are necessary to constrain the filling factor regions and infer plausible values. Even though we expect the extrapolations to be neither the most accurate nor mathematically flawless, our estimates try to be as reasonable as possible. Further work, aimed at an improvement in the filling factor data sets, would most likely lead to better estimates.

Finally, our simulations allow the spots to be placed randomly on the stellar surface and use a lognormal spot size distribution (\citealt{Korhonen2015} and references therein) peaking at 1$^{\circ}$ in radius. This choice is motivated by the fact that larger spots are more easily modelled via tomographic imaging, so their effects can be efficiently filtered out, even though the spot jitter increases with the spot size.

\subsection{Stellar rotation period}\label{sec:Prot}

We compute the stellar rotation period ($\mathrm{P}_{\mathrm{rot}}$) employing activity-rotation relations available in the literature. Either derived using chromospheric activity \citep{Noyes1984,SuarezMascareno2018} or coronal activity indexes \citep{West2008,Wright2013}, the relations show that the activity level increases with decreasing rotation period until a saturation regime is reached.

For M type stars, we use the formula provided by \citet{SuarezMascareno2018}, whereas for F, G, and K type stars, we use the formalism given by \citet{Noyes1984}. They obtained an empirical fit relating the magnetic activity to the Rossby number $\mathrm{Ro}=P_{\mathrm{rot}}/\tau_{\mathrm{conv}}$, where the convective turnover time ($\tau_{\mathrm{conv}}$) is a function of spectral type. Therefore, inverting the fit yields an expression for $P_{\mathrm{rot}}$ as function of $\log R'_{\mathrm{HK}}$ and $B-V$ colour index. 

The computed periods for our sample are between 8 and 54 days (see Table~\ref{tab:sim_params}), hence falling in the non-saturated regime of the activity-rotation relation.

Our code computes the projected rotational velocity of the star (vsini) combining the rotation period and the stellar radius (obtained from the stellar mass, see \citealt{Korhonen2015}). Because we assume inclination of 90$^\circ$ (see Section~\ref{sec:scalings}), these values are equivalent to the equatorial velocity. We report the estimates in Table~\ref{tab:sim_params} for completeness. Note that the values of both $\mathrm{P}_{\mathrm{rot}}$ and vsini are estimates from empirical relations, and as such cannot be considered  to be analogous to the ones measured from observations. We also note that for some of the stars in our sample there are estimates of the vsini and rotation period, but not for all, and the reported values are not always consistent.

\subsection{Active latitudes}\label{sec:latitudes}

There is evidence that the latitude at which spots emerge shifts toward the poles as the rotation of the star increases \citep{Schuessler1996,Granzer2000,Isik2018}. \citet{Schuessler1992} suggested that the Coriolis force affects the motion of the flux tubes as they rise to the surface: when the Coriolis force exceeds the buoyancy force, the flux tubes ascend parallel to the rotation axis and the star shows polar spots as a consequence.

\citet{Granzer2000} shows distributions of latitude emergence for rotation rates in range $0.25\mathrm{-}63\;\Omega_{\bigodot}$ or, equivalently, $P_{\mathrm{rot}}\simeq 0.4\mathrm{-}112\;\mathrm{days}$, and for different stellar masses. In our simulation sample, the fastest rotator is HD\,179949 with a period of 8 days, i.e. $\Omega=3.5\;\Omega_{\bigodot}$, and the slowest rotator is GJ\,180 with a period of 54 days, i.e. $\Omega=0.52\;\Omega_{\bigodot}$. Considering both the stellar mass and the rotation period of our stars, we use the latitude distributions in \citet{Granzer2000} to infer values of the active latitude bands for our simulation sample (see Table~\ref{tab:sim_params}).

%% file: Sec4.tex
\section{Simulations} \label{sec:simulations}
\subsection{Setup}
Using the collected parameters (Table~\ref{tab:sim_params}), we compute stellar surface maps with \textsc{spotss} and the associated synthetic spectra with \textsc{deema}. A detailed explanation of these codes is found in \citet{Korhonen2015}. 

\begin{figure}[t]
    \centering
    \includegraphics[width=0.7\columnwidth]{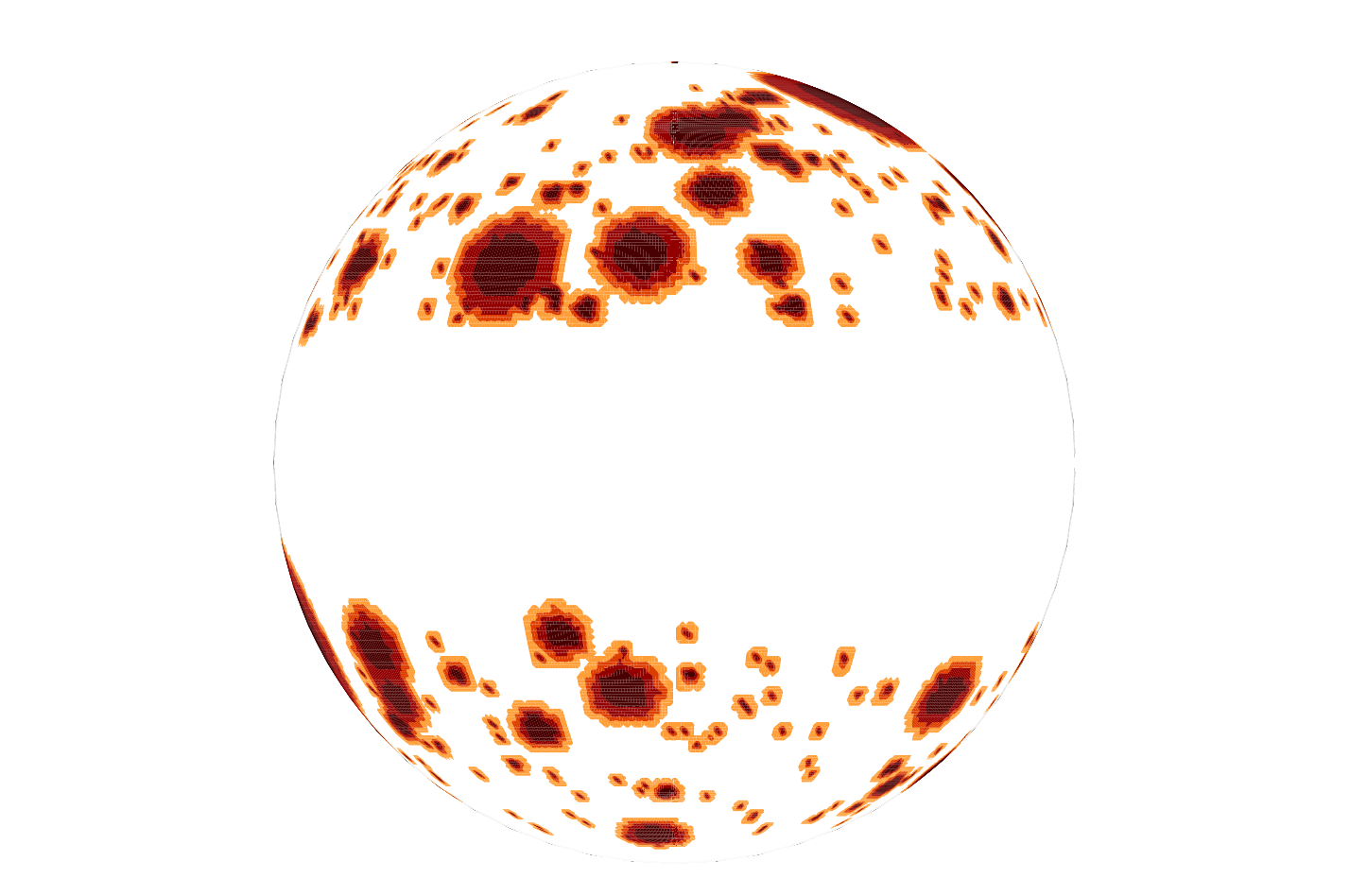}
    \includegraphics[width=\columnwidth]{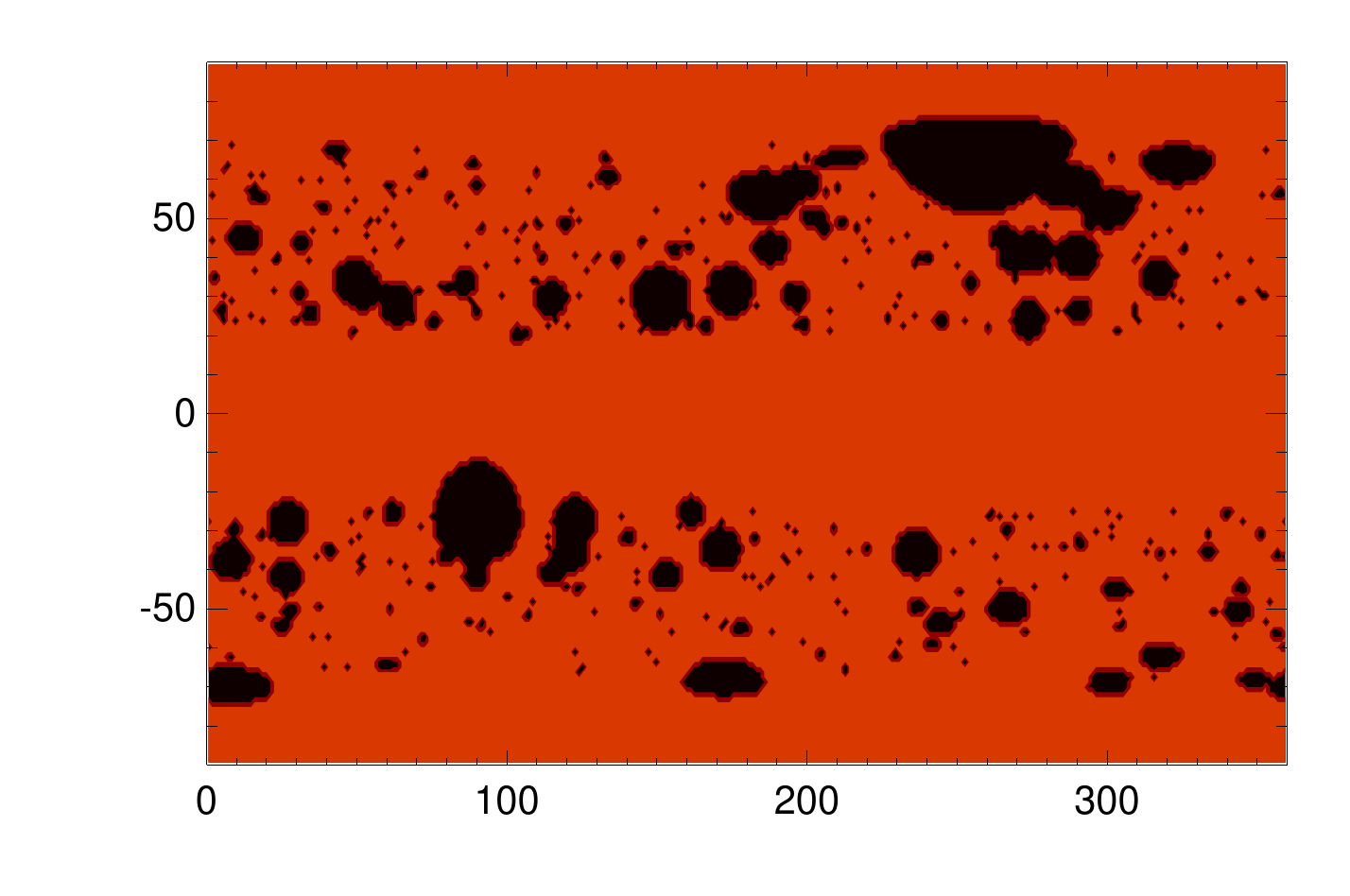}
    \caption{Top: Example of stellar surface map for $\varepsilon$\,Eri seen equator-on. Bottom: Mercator projection of the stellar surface. The $x$ and $y$ axes represent latitude and longitude, respectively, and are expressed in degrees. These maps are generated with $T_{\text{eff}}=5100\;\text{K}$, $T_{\text{spot}}=3950\;\text{K}$, $T_{\text{penumbra}}=4545\;\text{K}$, ff$= 20\;\%$ and $\Delta l = \pm(20-70)^{\circ}$. \label{fig:starmap}}
\end{figure}

Each surface map represents a certain spot configuration (including umbra and penumbra), which is assumed to change at every stellar rotation. Fig.~\ref{fig:starmap} illustrates an example for $\varepsilon$\,Eri. A map is accompanied by a number of synthetic spectra corresponding to the number of rotation phases. 

\begin{figure}[t]
    \centerline{\includegraphics[width=\columnwidth]{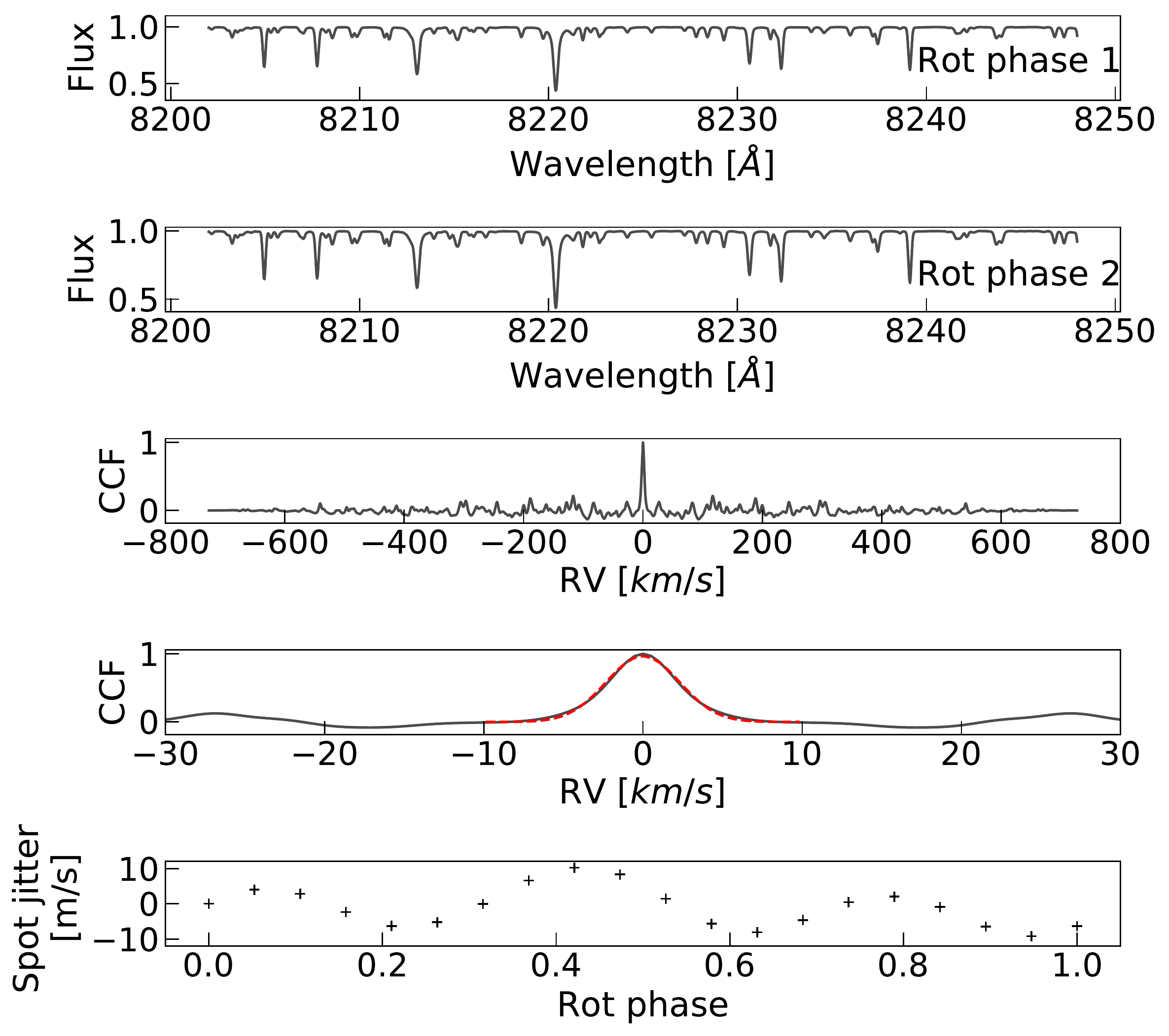}}
    \caption{From the top: synthetic spectrum of the first rotation phase, synthetic spectrum of the second rotation phase, normalized CCF between the first and second rotation phase spectra, Gaussian fit of the CCF around the peak, spot-induced jitter for every rotation phase. In the bottom plot, the values are vertically shifted so that the jitter at the first rotation phase is zero. This example is for $\varepsilon$\,Eri, the most active K type star in our sample. \label{fig:spot_jitter}}
\end{figure}

The RV curve is obtained as illustrated in Fig.~\ref{fig:spot_jitter}. The synthetic spectra are cross-correlated using the spectrum corresponding to the first rotation phase as a template. The choice of which rotation phase spectrum to take as template is rather arbitrary since it does not alter the results. We perform a Levenberg-Marquardt least squares fit of the maximum of the normalized cross correlation function (CCF) within $\pm10$ km s$^{-1}$ and using a Gaussian function, in order to find the value of the RV jitter. 

\begin{figure}[t]
    \centerline{\includegraphics[width=\columnwidth]{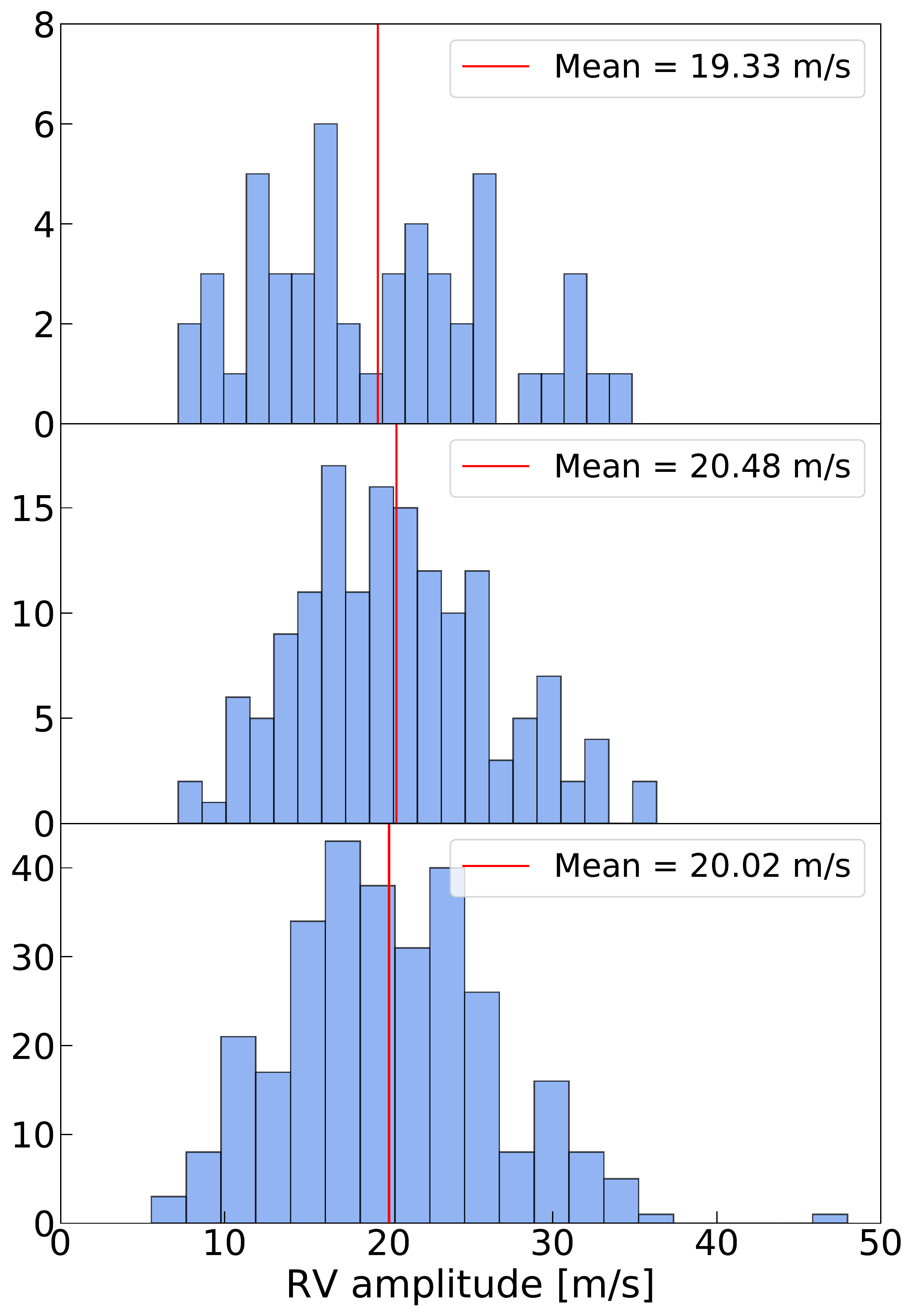}}
    \caption{Test to determine the optimal number of surface maps to simulate, taking $\varepsilon$\,Eri as example. We consider a grid of 50, 100, 150, 200, 250, and 300 random surface maps and compute the mean RV jitter peak-to-peak amplitude in each case. From top to bottom, the distributions of the RV jitter amplitude for the 50, 150, and 300 cases are shown. Using 150 maps results in a symmetric distribution of the RV jitter amplitude, with a $\leq$50 cm/s difference relative to the 300 case. \label{fig:nmaps_test}}
\end{figure}

We test the optimal number of simulated surface maps to ensure a reliable statistics of the jitter amplitude and a practical computation time. From Fig.~\ref{fig:nmaps_test}, we note that using 150 maps results in a reasonably symmetric distribution of the RV jitter amplitude, therefore satisfying our requirements. In addition, we set the number of synthetic spectra in each map equal to 20 to have a dense sampling of the RV curve.

\subsection{Scaling of the jitter}\label{sec:scalings}

\begin{figure}[t]
    \centerline{\includegraphics[width=\columnwidth]{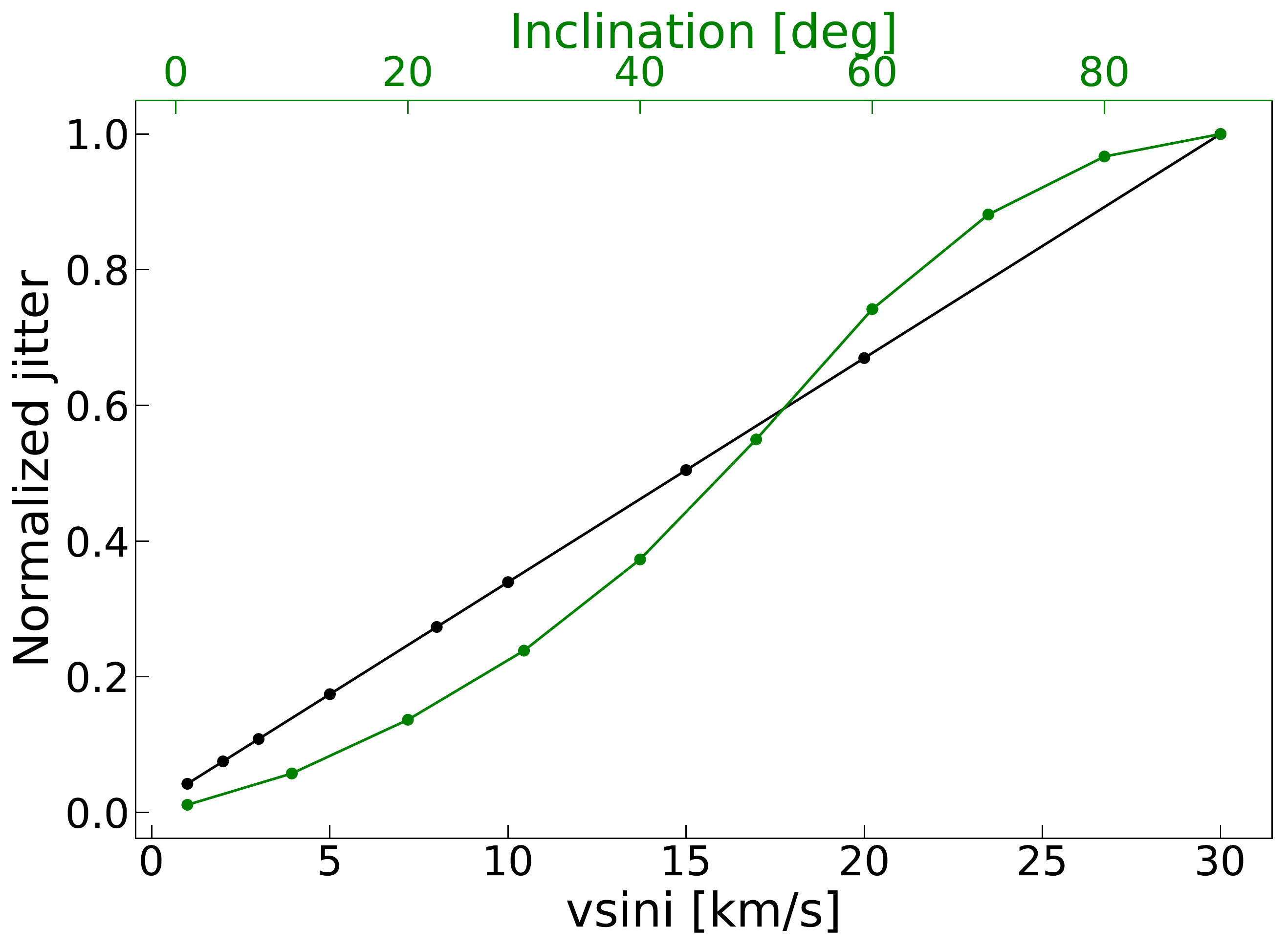}}
    \centerline{\includegraphics[width=\columnwidth]{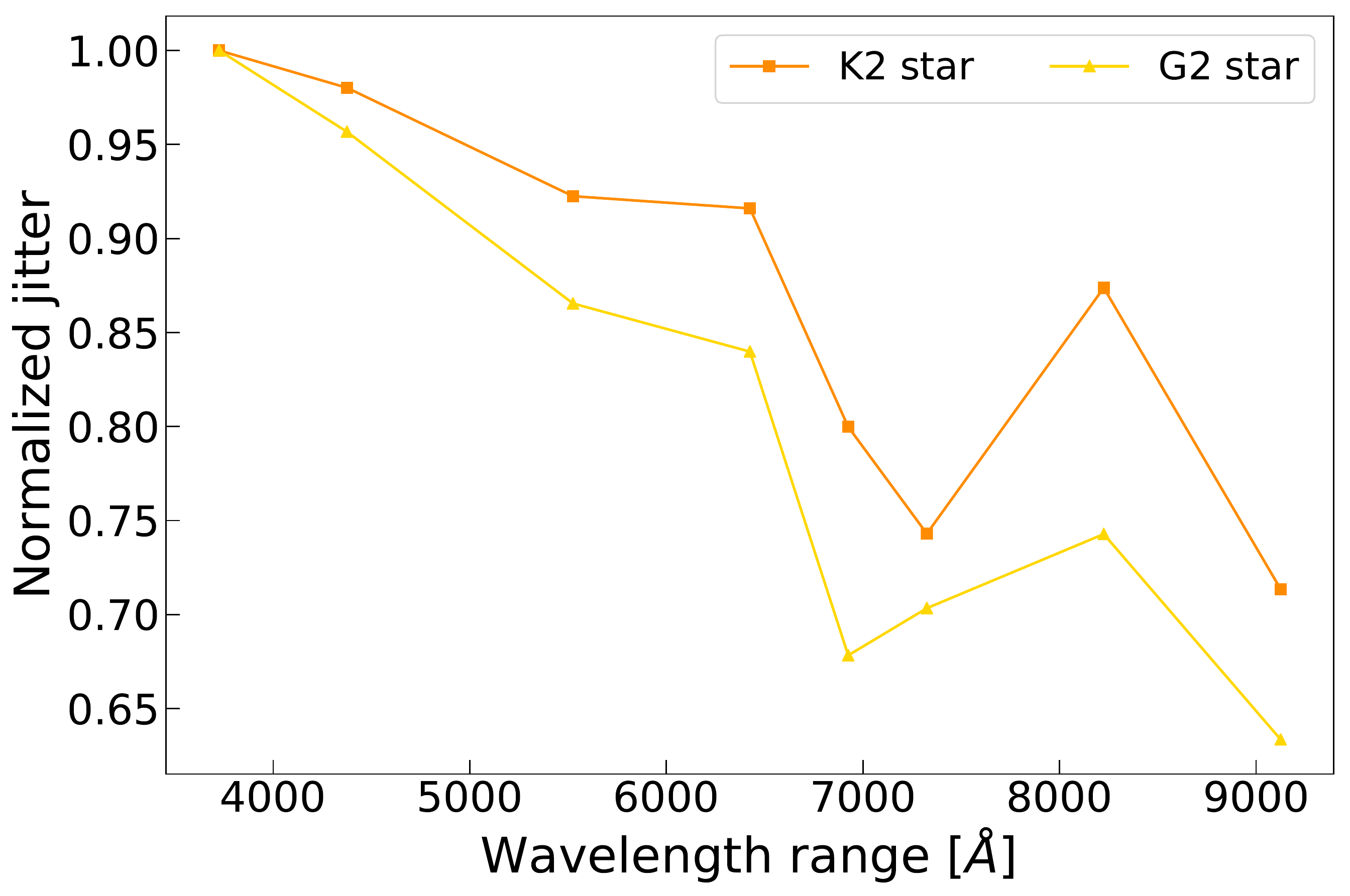}}
    \caption{
    Top: Scaling of the normalized jitter amplitude with vsini (black data points) and inclination (green data points). When the inclination effects are studied, a fixed value of vsini is assumed, and vice versa. The jitter increases as the view of the star is progressively equator-on and linearly with vsini. The values are reproduced from \citet{Korhonen2015}. Bottom: Scaling of the normalized jitter with spectral coverage for a G2 (yellow triangles) and a K2 (orange squares) type star. The values on abscissa are the averages of the 46\r{A} wide wavelength intervals examined. Bluer spectral regions are affected by a larger ($\sim30$\%) jitter amplitude. In all cases, the jitter amplitude is normalized to the largest value, and the simulated spot is equatorial.
    \label{fig:scalings}}
\end{figure}

We illustrate how our estimates of the activity jitter scale with parameters such as inclination, vsini and wavelength coverage of the spectrum. These dependencies, which we display in Fig.~\ref{fig:scalings}, have been previously investigated in several works \citep[e.g.,][]{Desort2007,Lagrange2010,Korhonen2015}.

For inclination and vsini, we reproduce the jitter values simulated by \citet{Korhonen2015} for a solar-like star (5800 K) and a 5$^\circ$ equatorial spot (4000 K). Changing the inclination from an equator-on view to the pole-on view leads to a reduced jitter, while stellar rotation correlates with the jitter. Note that for our simulation sample, we find an inclination of $60^{\circ}$ for GJ 3634 \citep{Bonfils2011} and $\sim90^{\circ}$ for HAT-P-5 and WASP-1 \citep{Bakos2007,Bakos2015,Simpson2011}. However, to ensure that the comparison of the results is consistent, we set the inclination $i$ to $90^{\circ}$ for all our simulated stars. 

For our simulations, we synthesize spectra covering between 8202 \r{A} and 8248 \r{A} and with an associated resolving power of $R=100,000$. Although the wavelength range is rather narrow, which is not the case for typical observations, no source of instrumental noise is included (ideal scenario). Thus, an extension of the wavelength interval would not reflect in an appreciable improvement of the CCF analysis. Moreover, \citet{Korhonen2015} showed that the jitter amplitude decreases when increasing the wavelength interval width up to 40 \r{A} and between 40--70 \r{A} the jitter stays virtually constant.

Note also that this wavelength region has smaller intrinsic jitter than many bluer regions \citep{Reiners2010}. This is illustrated in Fig.~\ref{fig:scalings} for the following wavelength intervals: [3700,3750], [4350,4400], [5500,5550], [6400,6450], [6900,6950], [7300,7350], [8200,8250], [9100,9150] \AA. We consider a G2 and a K2 type star with the temperature properties of the Sun and $\varepsilon$ Eri, respectively, and a single 1$^\circ$ radius equatorial spot, resolution of 100,000, inclination of 90$^\circ$ and vsini of 2.0 km s$^{-1}$. In this low activity scenario, the jitter amplitude ranges 70--44 cm s$^{-1}$ and 59--42 cm s$^{-1}$ for the G2 and K2 star, respectively. The interval of our simulations leads to a jitter amplitude that is 10\% and 25\% lower than the bluest one for the G2 and K2 star.

%% file: Sec5.tex
\section{Results and discussion}\label{sec:results}
From the RV curves of our simulations, we compute the mean peak-to-peak RV amplitude ($\langle \mathrm{RV}_{\mathrm{pp}} \rangle$) and its minimum ($\langle \mathrm{RV}_{\mathrm{pp}} \rangle_\mathrm{min}$) and maximum ($\langle \mathrm{RV}_{\mathrm{pp}} \rangle_\mathrm{max}$) values over all the spot configurations for each star. The \textsc{deema} code produces the V-band light curves associated with the spots configurations as well, therefore we include information on the photometric variability: minimum ($\Delta\mathrm{I}_\mathrm{min}$), maximum ($\Delta\mathrm{I}_\mathrm{max}$) and mean amplitude ($\langle\Delta\mathrm{I}\rangle$) of the light curve. This way we have a more complete view on the typical spot-driven jitter and photometric variability. The values are listed in Table~\ref{tab:sim_results}.  

\subsection{Consistency with empirical relations}

\begin{figure}[t]
    \centering
    \includegraphics[width=\columnwidth]{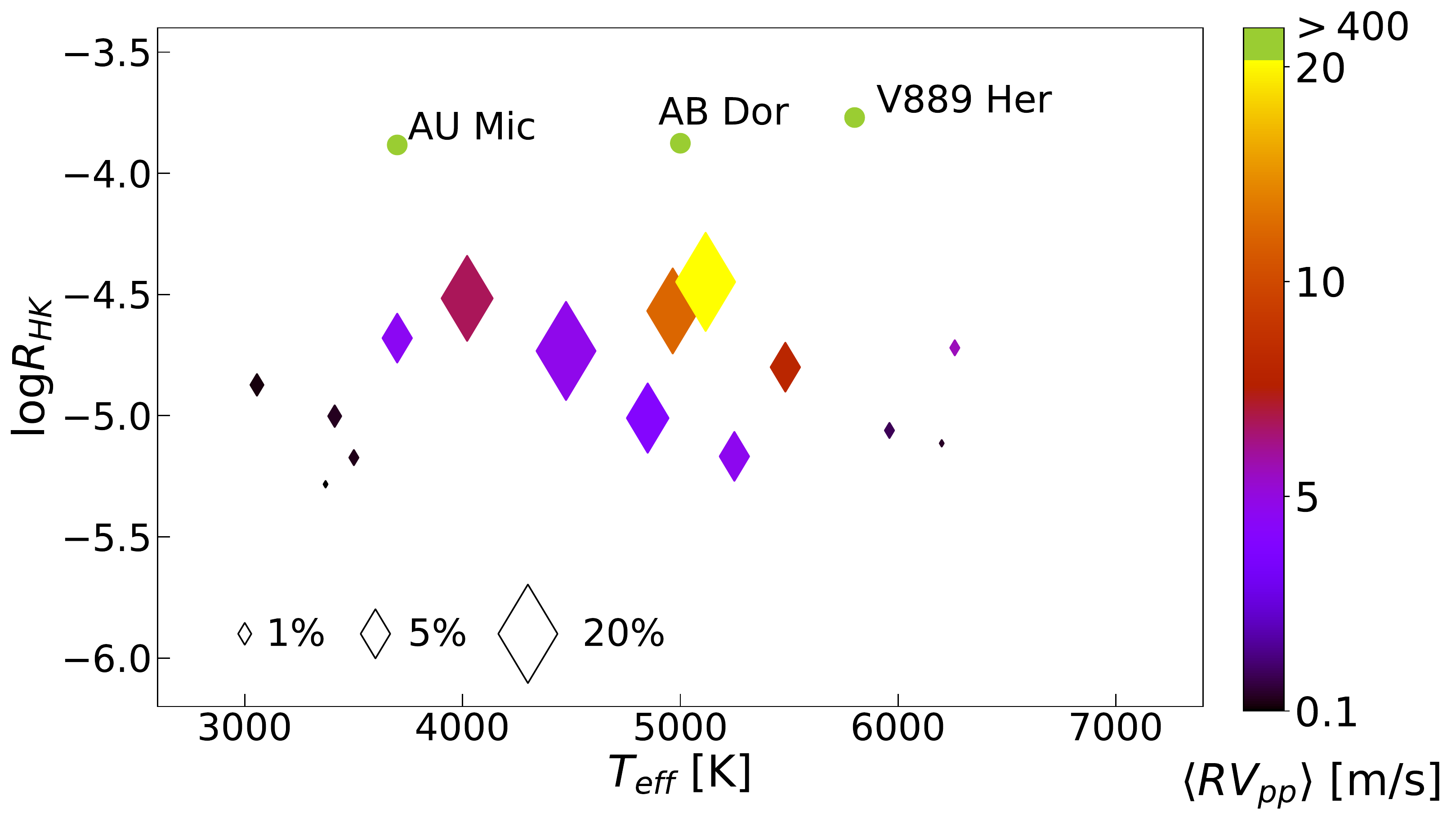}
    \caption{Stellar parameter space of our simulations and corresponding spot-induced RV jitter. The sample of 15 simulated host stars is shown, with data point size and colour encoding the spot filling factor and RV jitter amplitude, respectively. We include observational data (green) for three active stars (AU Mic, AB Dor and V889 Her) for a comparison of activity level and jitter with respect to our sample. The size of these three data points does not scale with the filling factor, as this information is not always available. \label{fig:results_plot}}
\end{figure}

Figure~\ref{fig:results_plot} illustrates the RV results as a function of the stellar parameters collected in Section~\ref{sec:simparameters}. The purpose of this plot is twofold: it allows a reasonable estimate of the RV jitter amplitude based on specific stellar parameters, hence informing observing strategies, and displays empirical relations between various quantities. Overall, we observe an increase of $\langle \mathrm{RV}_{\mathrm{pp}} \rangle$ with $\log R'_{\mathrm{HK}}$, which is expected by construction since higher activity translates in higher noise in RV data sets. This is consistent with the empirical relation between the RV dispersion and $\log R'_{\mathrm{HK}}$ described in \citet[e.g.,][]{Saar1998, Hojjatpanah2020}. As can be observed by the color gradient, $\langle \mathrm{RV}_{\mathrm{pp}} \rangle$ follows a parabolic trend over $T_{\mathrm{eff}}$ similarly to the filling factor (Fig. 10 in \citealt{Berdyugina2005}), which stresses the importance of knowing the filling factor and determining robust empirical relations involving it, as it is a crucial factor in regulating the jitter amplitude.

\begin{figure}[t]
    \centering
    \includegraphics[width=\columnwidth]{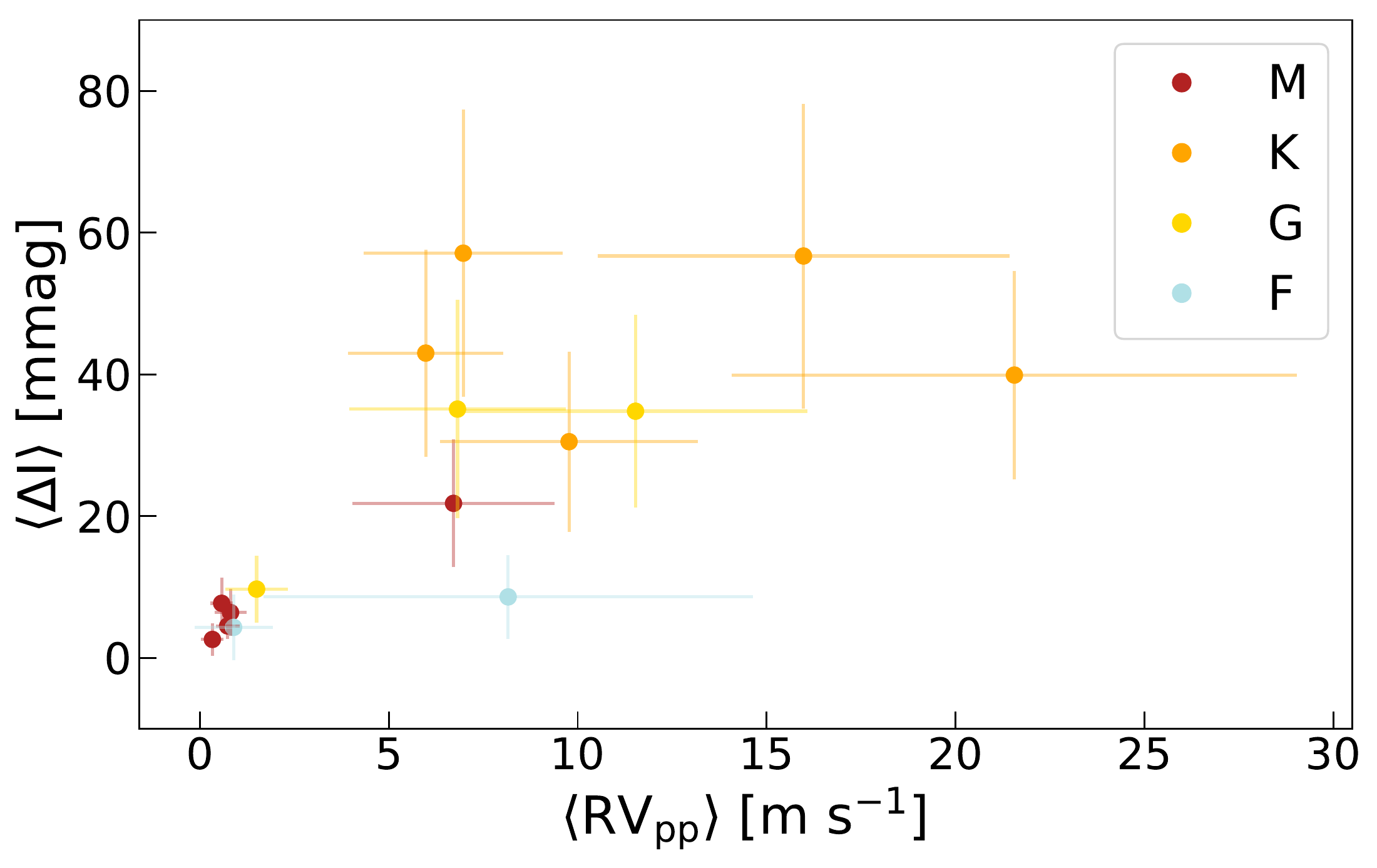}
    \caption{Correlation between the mean photometric amplitude and mean peak-to-peak RV jitter for the simulation sample. Different colors encode different spectral types. \label{fig:corr_plot}}
\end{figure}

The photometric amplitude shows a strong dependence of filling factor and spot-temperature contrast as well, in agreement with \citet{Johnson2021}. Moreover, the fact that photometric and jitter amplitude follow similar trends is an indicator of their correlation \citep{Hojjatpanah2020}. In Fig.\ref{fig:corr_plot}, we note that for the most active stars in our sample the clarity of the correlation diminishes (as indicated by the larger error bars as well). In particular, we observe that the largest $\langle \mathrm{RV}_\mathrm{pp}\rangle$ is 21.6 m s$^{-1}$ for $\varepsilon$ Eri (the most active star in our sample), while the mean photometric amplitude peaks at 57.1 mmag for HIP\,90979 (0.3 dex less active than $\varepsilon$ Eri), whose estimated jitter is 7.0 m s$^{-1}$. This can be explained by the different combination of the stellar parameters: the two stars share the same value of spot filling factor (20\%), but the spot-photosphere temperature contrast is 200 K lower for HIP\,90979, hence the spot configuration induces a smaller jitter \citep{Lagrange2010, Reiners2010}. At the same time, the estimated rotational velocity of $\varepsilon$ Eri is higher than for HIP\,90979, which leads to an enhanced jitter amplitude \citep{Desort2007,Korhonen2015}. Furthermore, the active latitudes of HIP\,90979 are $\sim$10$^\circ$ shifted towards the equator, implying a wider excursion of the spot impact on the light curve and therefore a larger photometric amplitude.

We observe a large jitter amplitude for HD\,179949, the earliest star in our sample, reaching a value similar to active K dwarfs (HIP\,54373, HIP\,90979). The most likely explanation is not by the correspondingly large value of spot filling factor or spot-photosphere temperature contrast (as WASP-1 features an analogous value), but in a fast rotation. In fact, HD\,179949  has the largest vsini in our sample (Table~\ref{tab:sim_params}).

With the maximum and minimum of both RV jitter and photometric amplitude, we note that the range of plausible values correlates with the activity level of the star, as expected. These ranges also reflect different spot configurations: few and clustered big spots on one side, and small numerous spots scattered over the surface on the other side. Taking GJ\,180 as an example, these two extremes likely correspond to the simulated maximum and minimum photometric amplitudes of 0.3 and 15.1 mmag, respectively.

\begin{center}
\begin{table*}[t]
    \caption{Results of the simulations: minimum, maximum and mean peak-to-peak RV amplitude, minimum, maximum and mean amplitude of the photometric variations. All quantities are computed using 150 different spot configurations and the error bars indicate the 1$\sigma$ uncertainty. \label{tab:sim_results}}
    \centering
    \begin{tabular*}{42pc}{@{\extracolsep\fill}lccrccr@{\extracolsep\fill}}
        \toprule
        \textbf{ID} & $\langle\mathrm{\textbf{RV}}_\mathrm{\textbf{pp}}\rangle_\mathrm{\bf min}$ [\textbf{m/s}] & $\langle\mathrm{\textbf{RV}}_\mathrm{\textbf{pp}}\rangle_\mathrm{\bf max}$ [\textbf{m/s}] & $\langle \mathrm{\textbf{RV}}_{\mathrm{\textbf{pp}}} \rangle$ [\textbf{m/s}] & $\Delta\mathrm{\bf I}_\mathrm{\textbf{min}}$ [\textbf{mmag}] & $\Delta\mathrm{\bf I}_\mathrm{\textbf{max}}$ [\textbf{mmag}] & \textbf{$\langle \Delta\mathrm{\textbf{I}} \rangle$} [\textbf{mmag}]\\
        \midrule
        YZ\,Cet & 0.2 & 2.0 & 0.6$\pm$0.3 & 2.4 & 24.4 & 7.7$\pm$3.6\\
        GJ\,180 & 0.04 & 1.8 & 0.3$\pm$0.3 & 0.3 & 15.1 & 2.6$\pm$2.3\\
        GJ\,687 & 0.2 & 2.8 & 0.8$\pm$0.4 & 1.6 & 20.9 & 6.4$\pm$3.3\\
        GJ\,3634 & 0.3 & 2.7 & 0.7$\pm$0.3 & 1.1 & 15.6 & 4.5$\pm$1.8\\
        GJ\,649  & 2.0 & 17.0 & 6.7$\pm$2.7 & 5.2 & 51.4 & 21.8$\pm$9.0\\
        HIP\,54373  & 4.0 & 19.7 & 9.8$\pm$3.4 & 10.9 & 63.2 & 30.5$\pm$12.7\\
        HIP\,90979 & 2.9 & 21.2 & 7.0$\pm$2.6 & 20.0 & 134.2 & 57.1$\pm$20.3\\
        HD\,156668 & 2.0 & 18.2 & 6.0$\pm$2.1 & 15.8 & 100.2 & 43.0$\pm$14.6\\
        HD\,192263 & 5.4 & 32.9 & 15.9$\pm$5.5 & 20.6 & 147.2 & 56.7$\pm$21.5\\
        $\varepsilon$\,Eri & 6.3 & 53.6 & 21.6$\pm$7.5 & 9.0 & 96.0 & 39.9$\pm$14.7\\
        HD\,11964  & 0.9 & 16.0 & 6.8$\pm$2.9 & 3.4 & 80.4 & 35.1$\pm$15.4\\
        HD\,49674 & 4.0 & 29.7 & 11.5$\pm$4.5 & 8.5 & 88.0 & 34.8$\pm$13.6\\
        HAT-P-5 & 0.4 & 5.1 & 1.5$\pm$0.8 & 2.1 & 30.6 & 9.7$\pm$4.7\\
        WASP-1 & 0.1 & 9.4 & 0.9$\pm$1.0 & 0.4 & 41.2 & 4.3$\pm$4.6\\
        HD\,179949 & 1.8 & 43.2 & 8.2$\pm$6.5 & 1.6 & 43.0 & 8.6$\pm$5.9\\
        \bottomrule
    \end{tabular*}
\end{table*}
\end{center}

\subsection{Comparison to observations}

We compare some of our spot-induced jitter amplitudes to observed values. For $\varepsilon$ Eri, our simulated value of 21.6$\pm$7.5 m s$^{-1}$ is comparable (within error bars) to the reported values of 25-30 m s$^{-1}$ \citep{Giguere2016,Petit2021}; for HD\,156668, we estimate 6.0$\pm$2.1 m s$^{-1}$, which is compatible with the 8 m s$^{-1}$ amplitude of the residuals of a Keplerian fit \citep{Howard2011}; for YZ Cet, an amplitude of 1.4 m s$^{-1}$ was obtained via Gaussian Process modelling \citep{2017bAstudillo}, which is larger than our predicted value by a factor of 2.3, but it is within the range of maximum and minimum amplitudes we computed for the star; for HD\,192263, we take the RV data set analyzed by \citet{Dragomir2012}, apply a Keplerian fit with the parameters indicated by the authors and find a $\sim$60 m s$^{-1}$ RV amplitude of the residuals, which is two and four times larger than our simulated maximum and mean peak-to-peak amplitude, respectively. 

Quantitative discrepancies between the simulated and observed values are expected since our computation considers starspots as the principal source of jitter, but other phenomena such as the inhibition of convective blueshift induced by faculae may dominate. Indeed, the impact of this effect becomes more relevant for G-type stars \citep{Meunier2010,Meunier2017,Miklos2020}, hence the derived amplitudes may be underestimated in these cases. In addition, the wavelength range we are considering is located on the red side (8200 -- 8245 \r{A}) of the optical spectrum for typical observations. Given that the jitter is expected to decrease with wavelength due to a lower contrast \citep{Desort2007,Reiners2010}, this may lead to underestimated values. Overall, the fact that some of our estimates are consistent with observations validates retroactively the assumptions on the stellar parameters (e.g. filling factor) we made in Section\ref{sec:simparameters}. 

We also investigate whether the choice of latitude bands has a significant impact on the jitter and photometric variability estimates. Indeed, restricting the spot appearance within latitude bands increases their uniformity, thus leading to a reduced jitter. We simulate the spotted surface of GJ\;649 (our most active M dwarf) without latitude constraints and compare the new estimates. We find $\langle\mathrm{RV}_\mathrm{pp}\rangle_\mathrm{min}$= 2.4 m s$^{-1}$, $\langle\mathrm{RV}_\mathrm{pp}\rangle_\mathrm{max}$= 16.5 m s$^{-1}$, and $\langle \mathrm{RV}_\mathrm{pp} \rangle$ = 7.9$\pm$3.2 m s$^{-1}$, which are only slightly larger ($<$1 m s$^{-1}$) than the values computed with the latitude bands. Likewise, we obtain $\Delta\mathrm{I}_\mathrm{min}$= 3.8 mmag, $\Delta\mathrm{I}_\mathrm{max}$= 47.2 mmag, and $\langle\Delta\mathrm{I}\rangle$= 23.9$\pm$9.6 mmag, which are at most 4 mmag greater than the constrained latitudes case. Therefore, for both jitter and photometric amplitudes and for a given star, we observe only a marginal effect of latitude bands.

In Figure~\ref{fig:results_plot}, we also locate the position of AU Mic, AB Dor and V889 Her. These are three very active stars, whose reported $\log R'_{\mathrm{HK}}$ is between -3.8 and -3.9 \citep{Strassmeier2000,Saikia2018}, i.e. 0.7 dex greater than the most active star in our sample. Such values reflect in a jitter amplitude of 400 m s$^{-1}$ for AU Mic \citep{Plavchan2020} and 600 m s$^{-1}$ for AB Dor and V889 Her \citep{Jeffers2014,Korhonen2015}, at least one order of magnitude greater than for $\varepsilon$ Eri. This approximate upper limit defined by AU Mic, AB Dor and V889 Her gives a better sense of the low and moderate activity level of our simulation sample and helps locating a preferential region in which the spot-induced jitter is below 50 m s$^{-1}$.

\section{Conclusions}

In this paper, we have carried out simulations of the spotted surface for a sample of 15 known host stars representative of the M--F spectral type range, and provided their typical spot-induced jitter amplitudes (see Fig.~\ref{fig:results_plot}). The peak-to-peak RV jitter in the visible domain for stars with similar properties can then be estimated from these reference values.

Prior information on the activity level was collected from the literature or measured by us: the simulation sample is characterized by $\log R'_{\mathrm{HK}}$ intermediate between -4.5 and -5.3, i.e. belonging to the moderatively active and inactive regime. This range originates directly from the fact that the planet searches have primarily targeted inactive stars. The activity index was used to characterise the starspot activity on these stars as accurately as possible. Finally, the stellar parameters required to simulate the spotted surfaces were also retrieved from other works or estimated from empirical relations. 

We can draw the following conclusions from our simulations:
\begin{itemize}
    \item We observe a positive correlation between the RV and photometric amplitude, as well as with $\log R'_{\mathrm{HK}}$, in accordance with \citep{Hojjatpanah2020}. The departure from a tight correlation for some stars can be explained by their different activity levels and by a certain combination of stellar properties, consistently with known empirical trends (e.g. increased jitter with vsini or spot-photosphere temperature contrast).
    \item When compared to observational values, our estimates of $\mathrm{RV}_\mathrm{pp}$ show a reasonable agreement. Discrepancies are a direct consequence of our stellar parameter approximations and of our limited model. Knowledge about, e.g., the spot filling factor is scarce and restricted to more active stars than our simulation sample, hence our estimates rely inexorably on extrapolations. At the same time, the exclusion of faculae (and their effect on the suppression of convective blueshift) may contribute to the discrepancy of few m/s \citep{Milbourne2019}.
    \item Even for the low activity levels characterizing our sample, the RV signal induced by spots only is once again demonstrated to be a nuisance for small planet searches. It is indeed comparable to or greater than the signature induced by an Earth-mass planet, which is on the order of cm s$^{-1}$ and m s$^{-1}$ for G- and M-type stars, respectively. This emphasizes the importance of an appropriate target selection for RV surveys aimed at finding Earth's siblings. In this sense, Fig.~\ref{fig:results_plot} can be used as reference to estimate the expected $\mathrm{RV}_\mathrm{pp}$ of targets for RV searches.
\end{itemize}

In this work we have provided means to characterise starspot properties for moderately active and inactive stars spanning spectral types M--F. We also provide estimates of the jitter caused by these starspot to help plan future planet searches using radial velocity method. Additional constraints of the stellar surfaces over different spectral types from observations are required to proceed further in this work. In fact, our simulations would benefit from more precise star spot properties and from the additional modelling of stellar faculae, overall leading to more realistic estimates of the activity jitter.

%% file: appendix.tex
\section{Stellar data from observations}

\begin{center}
\begin{table*}[t]
    \caption{Activity levels for main sequence and giant stars used as references in Fig.~\ref{fig:sample_ff}. All values of surface temperature and filling factor are taken from \citet{Berdyugina2005} and \citet{Andersen2015}, unless specified differently. Multiple entries for the same star correspond to different techniques with which the filling factor was measured. The $\log R'_{\mathrm{HK}}$ index is extracted from the indicated reference or computed by us when only the $S$ index is provided (similarly to Section~\ref{sec:Smeasures}). \label{tab:ff_data}}
    \centering
    \begin{tabular*}{42pc}{@{\extracolsep\fill}llcrl@{\extracolsep\fill}}
        \toprule
        \textbf{ID} & $\textbf{T}_{\mathrm{\textbf{eff}}}\;[\mathrm{\textbf{K}}]$ & \textbf{Spectral type} & \textbf{ff [\%]} & \textbf{log} $\textbf{R}'_{\mathrm{\textbf{HK}}}$\\
        \midrule
         &  & Giants &  & \\
        \midrule
        HD\,199178 & 5350 & G5 III & 29 & -4.056\tnote{$^m$}\\
        $\lambda$\,And & 4750 & G8 III & 23 & -4.476\tnote{$^i$}\\
        $\lambda$\,And & 4780 & G8 III & 17 & -4.476\\
        $\sigma$\,Gem & 4600 & K1 III & 33 & -4.360\tnote{$^n$}\\
        $\sigma$\,Gem & 4440 & K1 III & 8 & -4.360\\
        HR\,1099 & 4700 & K1 IV & 40 & -3.841\tnote{$^j$}\\
        IM\,Peg & 4450 & K2 III & 20 & -4.208\tnote{$^k$}\\
        IM\,Peg & 4400 & K2 III & 15 & -4.208\\
        IM\,Peg & 4666 & K2 III & 11 & -4.208\\
        IM\,Peg & 4666 & K2 III & 12 & -4.208\\
        IM\,Peg & 4666 & K2 III & 32 & -4.208\\
        IM\,Peg & 4440 & K2 III & 7 & -4.208\\
        II\,Peg & 4600 & K2 IV & 37 & -3.884\tnote{$^o$}\\
        II\,Peg & 4750 & K2 IV & 50 & -3.884\\
        II\,Peg & 4750 & K2 IV & 43 & -3.884\\
        II\,Peg & 4600 & K2 IV & 20 & -3.884\\
        IN\,Vir & 4600 & K2 IV & 40 & -4.397\tnote{$^p$}\\
        VY\,Ari & 4916 & K3 IV& 41 & -3.894\tnote{$^l$}\\
        VY\,Ari & 4600 & K3 IV& 15 & -3.894\\
        VY\,Ari & 4600 & K3 IV& 12 & -3.894\\
        VY\,Ari & 4600 & K3 IV& 15 & -3.894\\
        VY\,Ari & 4916 & K3 IV& 15 & -3.894\\
        VY\,Ari & 4916 & K3 IV& 16 & -3.894\\
        \midrule
         &  & Main sequence &  & \\   
        \midrule
        Sun min & 5870 & G2 V & 0.02 & -4.981\tnote{$^a$}\\
        Sun max & 5870 & G2 V & 0.50 & -4.931\tnote{$^a$}\\
        EK\,Dra & 5930 & G2 V & 6 & -4.103\tnote{$^b$}\\
        EK\,Dra & 5850 & G2 V & 11 & -4.103\tnote{$^b$}\\
        EK\,Dra & 5830 & G2 V & 40 & -4.103\\
        HD\,130322\tnote{$^d$} & 5330 & K0 V & 1\tnote{$^e$} & -4.552\\
        AB\,Dor & 5200 & K0 V & 5 & -3.877\tnote{$^b$}\\
        AB\,Dor & 5200 & K0 V & 12 & -3.877\\
        LQ\,Hya & 5175 & K2 V & 45 & -3.967\tnote{$^b$}\\
        OU\,Gem & 4925 & K3 V& 35 & -4.490\tnote{$^b$}\\
        V833\,Tau & 4500 & K4 V& 45 & -4.060\tnote{$^c$}\\
        EQ\,Vir & 4380 & K5 V & 45 & -3.864\tnote{$^b$}\\
        BY\,Dra & 4100 & M0 V& 34 & -3.811\tnote{$^b$}\\
        BY\,Dra & 4100 & M0 V& 60 & -3.811\\
        AU\,Mic & 3500 & M2 V & 10 & -3.883\tnote{$^b$}\\
        EV\,Lac & 3300 & M4 V & 7 & -3.749\tnote{$^b$}\\
        HU\,Del\tnote{$^f$} & 3200 & M4 V & 3\tnote{$^g$} & -4.498\tnote{$^h$}\\
        \bottomrule
    \end{tabular*}
    \begin{tablenotes}
    \item Sources: $^a$ \citet{Egeland2017}, $^b$ \citet{Saikia2018}, $^c$ \citet{Mishenina2012}, $^d$ \citet{Krejcova2012}, $^e$ \citet{Hinkel2015}, $^f$ \citet{Houdebine2019}, $^g$ \citet{Barnes2015}, $^h$ \citet{Houdebine2017}, $^i$ \citet{Gray2003}, $^j$ \citet{Gray2006}, $^k$ \citet{Isaacson2010}, $^l$ \citet{Wright2004}, $^m$ $S$ index from \citet{Duncan1991} and $B-V$ from \citet{Panov2007}, $^n$ $S$ index from \citet{Duncan1991} and $B-V$ from \citet{Ducati2002}, $^o$ $S$ index from \citet{Duncan1991} and $B-V$ from \citet{Rutten1987}, $^p$ $S$ index from \citet{Pace2013} and $B-V$ from \citet{Strassmeier2000}.
    \end{tablenotes}
\end{table*}
\end{center}